\documentclass[aps,prl,superscriptaddress]{revtex4}
\usepackage{graphicx}
\usepackage{rotating}

\begin{document}

\title{Deformed two center shell model}

\author{R. A. Gherghescu}
\email[]{rgherg@ifin.nipne.ro}
\affiliation{
National Institute of Physics and Nuclear
Engineering, RO-76900 Bucharest, Romania}

\date{\today}

\begin{abstract}
A highly specialized two-center shell model has been developed
accounting for the splitting of a deformed parent nucleus into two
ellipsoidaly deformed fragments. The potential is based on deformed
oscillator wells in direct correspondance with the shape change 
of the nuclear system. For the first time a potential responsible for 
the necking part between the fragments is introduced on potential
theory basis. As a direct consequence, spin-orbit {\bf ls} and 
{\bf l$^2$} operators are calculated as shape dependent. Level scheme
evolution along the fission path for pairs of ellipsoidaly 
deformed fragments is calculated. The Strutinsky method yields the
shell corrections for different mass asymmetries from the superheavy
nucleus $^{306}$122 and $^{252}$Cf all along the splitting process.

\end{abstract}


\maketitle

\section{Introduction}

From more than thirty years on, two-center shell models are precious
tools in the research of fusion, fission and cluster decay processes.
The capability of producing transition level schemes during the 
splitting process from an initial parent nucleus towards two different
fragments  offers the opportunity of studying what are the possible 
microscopic changes through a certain fission or decay channel.

In 1961 a symmetric double center oscillator potential has been analyzed
by Merzbacher \cite{mer}. A strong link was emphasized with regard to
diatomic molecules (like ammonia molecule) where the motion in the
neighbourhood of an equilibrium state is close to the one generated by a
harmonic potential. This problem has been solved for two identical 
spherical potential wells.

In the 1970's, J. Maruhn, W. Greiner and the Frankfurt school developed
the asymmetric two-center shell model \cite{mar}. It was an important step 
for the study of mass asymmetry in binary nuclear fission. The model 
considered two spherical asymmetric fragments subjected to two-oscillator
type potential. A remarcable feature of this model was its ability to describe
two overlapping spherical nuclei fission shapes up to their separated
configuration. Thus it was possible for microscopic effects to be held
responsible as being specifically due to the two emerging and gradually
separating level schemes. Basically it is under this form that two center
shell models are widely used today in nuclear fusion, fission and cluster
decay phenomena.

Two center states in light nuclei are taken into account in connection with
molecular orbitals in a two-center shell model picture \cite{oer}. Also,
application of the two-center shell model on the study of $^{17}O+^{12}C$
reaction has been published \cite{thi} with emphasis on nuclear Landau-Zenner
effect.

Nucleus-nucleus collisions have also been considered in the framework of the
two-center shell model to obtain the two-center level diagrams and single
particle corrections for asymmetric systems such as $^{16,17}O+^{24,25}Mg$
\cite{par}.

The importance of an adequate description of cold fission, cluster radioactivities
and alpha decay in terms of an asymmetric and {\it deformed} single particle
shell model with more realistic shapes during fission and fusion processes was
repeteadly stressed \cite{poe}.

Recently, attempts have been made to use the two-center shell model in synthesis
and decay of superheavy nuclei. It is a suitable theoretical model to study
the microscopic effects on possible projectile-target combinations in their way
from two different quantum systems to one. As the synthesized nucleus is heavier,
increasing Coulombian repulsion lowers the macroscopic potential barrier almost
to complete dissapearance. The only way a superheavy element can survive is due
to the shell effects \cite{ghe}. A two-center shell model is essential for the
description of fusion and fission of superheavy elements. It shows that the
shell structure of the two participating target-projectile nuclei is visible
far beyond the barrier into the fusioning nucleus and is crucial in choosing
the most favorable pair approaching each other through {\it cold} fusion
valleys. Transition behaviour of the two partner shells will provide shell
corrections which lower the fusion barrier to be overcome, as compared to  
neighbouring projectile-target combinations.

Up to now, all the variants of two-center shell models used spherical nuclei.
I shall consider now the motivation of this work. In any process which implies
a pass from two quantum systems to one (fusion) or the other way around (fission)
there are certain situations when one or both fission fragments or fusion partners
are deformed. Such a reaction could yield cold energy valleys due to deformed
shell structure of the participants. Fragment deformations are properly accounted
in this work. A deformed two-center shell model (DTCSM) is proposed, where the
main part of the potential consists of two ellipsoidaly deformed Nilsson type
oscillators for axially symmetric shapes. Any change in the nuclear surface shape
is reflected in a corresponding modification of the four oscillator frequencies
along the symmetry axis and perpendicular to it.

It is also well established that, especially in fission and cluster decay, a
necking region builds its way between the fragments, smoothly linking the
two ellipsoids one to the other. This work associates for the first time
a microscopic potential to a spherically matching neck region of the nuclear
shape. The neck potential is constructed in such a way that it takes the same
value on the necking region surface as on the ellipsoidal region surfaces
of the fragments. Equipotentiality is thus respected on the nuclear surface.

The usual spin-orbit and squared angular momentum operators are calculated
with help of the potential-dependent formulae ${\bf ls}=(\nabla V\ \times
{\bf p}){\bf s}$, and ${\bf l}^2=(\nabla V\ \times {\bf p})^2 $ \cite{rij}. 
The potential in this model follows exactly the nuclear shape, and so do
the  ${\bf ls}$ and  ${\bf l}^2$ operators. Finally, the levels diagram with
respect to the elongation, neck parameter and fragment deformations are used
to calculate the shell correction by means of the Strutinsky method.

Calculations are presented as an example for different mass asymmetries, 
for the superheavy nucleus $^{306}$122 and for $^{252}$Cf. The dependence 
on various neck parameter values and on different pairs of fragments is
discussed.

\section{Shapes}

Fig. 1 shows the main geometrical parameters defining
the axially symmetric shape family DTCSM is dealing with.
Two ellipsoids (the deformed fragments) with semiaxes $a_1$, $b_1$
and $a_2$, $b_2$ are, at a certain moment, separated at a distance
$R$ between the two centers $O_1$ and $O_2$. A sphere centered in
$O_3$ with radius $R_3$ is rolling around the symmetry axis, being
tangent all the time to the two ellipsoids. The necking region,
between the two tangent points, is generated in this way. Thus we have
five independent parameters to design the deformation space: two
fragment shape asymmetries $\chi _1=b_1/a_1$, $\chi _2=b_2/a_2$ 
(if $a_1$ and $a_2$ are given as the correspondent of $\beta _2$
for every $A_1$ and $A_2$, the other semiaxes are calculated from the total
volume conservation condition), mass asymmetry $A_1/A_2$, the neck
radius $R_3$ and the distance between centers $R$. Obviously, this
set is available for every parent nucleus $A,Z$ with its own
$\chi =b/a$.

A few shape sequences obtained by varying two of the parameters, $R_3$
and $R$ for the same parent and the same mass asymmetry $A_1/A_2$ are depicted
in Fig. 2. Every row starts with the same ellipsoidal parent on the left
hand side of the figure. As the distance between centers increases, the two
deformed fragments with fixed $\chi _1$ and $\chi _2$ shape asymmetries
begin to separate one from the other. Variation with the neck radius $R_3$
is noticeable on vertical direction. Shape sequences with very small necking 
region are shown in the upper row (for zero neck radius, $R_3=0$ fm, we get
compact shapes suitable for fusion reactions), passing through intermediary
neck radii, comparable to the magnitude of the ellipsoids semiaxes (second
and third row), down to the last row where large neck radius generates very
elongated shapes.

This kind of configurations will be microscopically treated along the variation
of the deformation space parameters.

\section{The potential}

The equations for shape surfaces described in the previous section
can be written in cylindrical coordinates (due to axyal symmetry) as:

\begin{equation}
\rho (z)= \left \{ \begin{array}{ccc}
\rho _1(z)= & [b_1^2- \chi _1 z^2]^{1/2} & -a_1 \leq z \leq z_{c1} \\
\rho _g(z)= &  \rho _3- [R_3^2-(z-z_3)^2]^{1/2}, & z_{c1} \leq z \leq z_{c2} \\
\rho _2(z)= &[b_2^2 - \chi _2(z-R)^2]^{1/2}, & z_{c2} \leq z \leq R+a_2
\end{array} \right.
\end{equation}
where the origin is placed in the center of the heavy fragment $O_1$. Neck sphere
center coordinates are $(z_3, \rho _3 )$, and $z_{c1}$ and $z_{c2}$ are the two
tangent points of the neck sphere with the two ellipsoids.

The oscillator potential correspondig to these two-center shapes must have the
same value on the nuclear surface. For spheres , for example we have:

\begin{equation}
V_0=\frac{m_0 \omega _i^2 R_i^2}{2}
\end{equation}
where $R_i$ is the radii of a nucleus with atomic mass $A_i$. Since 
$\hbar \omega _i=41A^{-1/3}$ and $R_i=r_0A^{1/3}$ (where $r_0\approx $ 1.16) then
$V_0 \approx 54.5 $ MeV. If we write these simple relations for the surface
of ellipsoidal shapes:

\begin{eqnarray}
 \frac{1}{2}m_0 \omega _{z_i}^2a_i^2= & V_0  \nonumber \\
 \frac{1}{2}m_0 \omega _{\rho _i}^2b_i^2= & V_0 
\end{eqnarray}
the frequencies $\omega _{z_i}$, $\omega _{\rho _i}$ are defined along 
the symmetry axis and respectively perpendicular to it, as functions of the two
ellipse semiaxes.

For an arbitrary origin, placed on the symmetry axis, the ellipsoids surface equations
read:

\begin{eqnarray}
\frac{\rho ^2}{
\displaystyle \frac{2V_0}{m_0 \omega _{\rho _1 }^2}}+
\frac{(z+z_1)^2 }{
\displaystyle \frac{2V_0}{m_0 \omega _{z_1}^2}}=1 \nonumber \\
\frac{\rho ^2}{
\displaystyle \frac{2V_0}{m_0 \omega _{\rho _2 }^2}}+
\frac{(z-z_2)^2 }{
\displaystyle \frac{2V_0}{m_0 \omega _{z_2}^2}}=1 
\end{eqnarray}
where $z_1$ and $z_2$ are the absolute values of each of the two centers
coordinates. Now the two oscillator potential expressions for deformed
fragments come straightforward:

\begin{eqnarray}
V_1(\rho ,z)=\frac{1}{2}m_0 \omega _{\rho _1}^2 \rho ^2+
\frac{1}{2}m_0 \omega _{z_2}^2(z+z_1)^2 \nonumber \\
V_2(\rho ,z)=\frac{1}{2}m_0 \omega _{\rho _2}^2 \rho ^2+
\frac{1}{2}m_0 \omega _{z_2}^2(z-z_2)^2
\end{eqnarray}

What we have left to establish is the necking region potential, $V_g(\rho,z)$.
The force that keeps nucleons confined within the ellipsoid is an attractive
type one, radially inward ($\vec{F}_{ellipsoid} \sim -\vec{r}$). The same reason leads
us to the hypothesis that, if nucleons are confined within the concave necking region,
which is geometrically {\it inversed} to the ellipsoid convex surface with respect to
the centers of the fragments, a rejective force is needed, radially outward:

\begin{equation}
\vec{F}(\vec{r})=- \beta \vec{r}
\end{equation}
hence a force oriented from outside the nuclear shape toward the surface.
Then the corresponding potential is related to the expression:

\begin{equation}
\int _0^r \vec{F}d\vec{r}=\varphi (r)-\varphi(0)
\end{equation}
or

\begin{equation}
\int _0^r \vec{F}d\vec{r}=\int _0^r - \beta \vec{r}d\vec{r}=
- \frac{\beta r^2}{2}
\end{equation}   
Consequently, $\varphi (r)=-\frac{\beta r^2}{2}$ is a potential form
generating the rejective force. In this way, the rejective neck potential,
defined up to a constant, must look like:

\begin{equation}
V_{g1}(r)=V_c+\varphi (r)
\end{equation}
where $V_c$ is a constant to be determined. On the nuclear surface $S_g$, we have
again:

\begin{equation}
V_{g1} \mid _{S_g}=V_c+\varphi (r) \mid _{S_g}=V_0
\end{equation}
or

\begin{equation}
V_c=V_0-\varphi (r) \mid _{S_g}
\end{equation}
For the sake of consistency, the rejective force is also considered of oscillator
type:

\begin{equation}
\vec{F}(\vec{r})=-m_0 \omega _g ^2 \vec{r}
\end{equation}
therefore
\begin{equation}
\varphi (r)=-\frac{m_0 \omega _g^2r^2}{2}
\end{equation}
where the frequency $\omega _g$ has to be found. Since the potential must follow
the geometrical shape, at the neck region the function $\varphi (r)=\varphi (\rho ,z)$
reads:

\begin{equation}
\varphi (\rho ,z)=-\frac{m_0 \omega _g^2}{2}
[(\rho - \rho _3)^2 +(z-z_3)^2]
\end{equation}
and is centered in the middle of the neck sphere $O_3(\rho _3,z_3)$. On the neck
surface then, where $(\rho ,z) \in S_g$, we have:

\begin{equation}
\varphi (\rho ,z) \mid _{S_g}=\frac{m_0 \omega _g^2}{2}R_3^2
=V_{g1} \mid _{S_g}=V_0
\label{eq:fi}
\end{equation}
Then $V_c=V_0- \varphi (r) \mid _{S_g}=2V_0$ and the total neck potential from
outside the shape down to the surface is:

\begin{equation}
V_{g1}(r)=2V_0-\frac{m_0 \omega _g^2}{2}[(\rho -\rho _3)^2+
(z-z_3)^2]
\end{equation}
and the neck frequency is directly related to the neck radius by Eq. \ref{eq:fi}.
$V_{g1}$ reach its maximum at the center of the neck sphere ($\rho =\rho _3,
z=z_3$), where $V_{g1}=2V_0$, then is decreasing down to the surface value
$V_{g1}=V_0$ at a distance equal to the neck radius
$(\rho - \rho _3)^2+(z-z_3)^2=R_3^2 $ from $O_3$.

To complete the neck-dependent potential, there still remains the region
inside the nuclear shape between the necking surface and the interior
contours of the two ellipsoids (black colored, denoted by $V_{g2}(\rho,z)$
in the upper part of Fig. 3). It can be observed that on the ellipsoids
surface {\it inside} the shape the deformed oscillator potential has the
same value as on the fragments surface, namely $V_0$. But on the surface of
the shape within the necking region, the value is also $V_0$. Then, one
concludes that inside the region volume between the neck surface and the two
ellipsoids surfaces - $\rho _1(z) \le \rho \le \rho _g(z)$ and
$\rho _2(z) \le \rho \le \rho _g(z)$ - the neck potential is constant.

\begin{equation}
V_{g2}(\rho ,z)=cst=V_0
\end{equation}

Finally, the deformed oscillator potential part for the DTCSM reads:
\begin{equation} 
V_{DTCSM}(\rho ,z)= \left \{ \begin{array}{ccc}
V _1(\rho ,z)= & \frac{1}{2}m_0 \omega _{\rho _1}^2 \rho ^2
+\frac{1}{2}m_0 \omega _{z1}^2(z+z_1)^2 & ,v_1 \\
V _g(\rho ,z)= &  \left \{ \begin{array}{ccc}
V_{g1}(\rho ,z)= & 2V_0- [\frac{1}{2}m_0 \omega _g^2(\rho - \rho _3)^2+
\frac{1}{2}m_0 \omega _g^2(z-z_3)^2] & ,v_{g1} \\
V_{g2}(\rho ,z)= & V_0 & ,v_{g2} 
\end{array} \right. \\
V_2(\rho ,z)= & \frac{1}{2}m_0\omega _{\rho _2}^2 \rho ^2+
\frac{1}{2}m_0 \omega _{z_2}^2(z-z_2)^2 & ,v_2
\end{array} \right.
\label{eq:os}
\end{equation}
where $v_1$, $v_{g1}$, $v_{g2}$ and $v_2$ are the spatial regions where the
corresponding potentials are acting. These regions are about to be revealed
further on.

\section{Matching potential surfaces}

One of the most important issues for the two-center shell models is that the
separation plane between the two fragments is not the one where the potentials
equal eachother. As an example, for compact shapes (no neck, $R_3=0$) the points
( $\{ \rho _s,z_s \} $ ) on the separation line (where the two ellipsae intersect)
result in different values for the potentials, $V_1(\rho _s ,z_s) \ne 
V_2(\rho _s ,z_s)$.
The sharp cusp between the two values makes calculation totally wrong. Attempts
to solve the problem by a geometrical transition function, making the frequency
$\omega _{\rho _1}(z)$ join $\omega _{\rho _2}(z)$ introduce an unpredictible
approximation which is bigger as the mass asymmetry is larger; moreover, it does 
not reproduce the neck spherical shape and still, only the $\rho -$part of the
potential is solved.

In conclusion, one has to divide the Hilbert space such that the {\it whole}
potential $V_{DTCSM}(\rho ,z)$ could smoothly and continuosly pass from one fragment 
to the neck and finally to the other fragment.

First, let us take the deformed fragment oscillators into consideration and find
out what are the spatial limits of continuity between the two potentials. The
equation for such a region must be complacent with the matching condition:

\begin{equation}
V_1(\rho ,z)=V_2(\rho ,z)
\end{equation}
where

\begin{equation}
\begin{array}{ccc}
V_1= & \frac{1}{2}m_0 \omega _{\rho _1}^2 \rho ^2+
\frac{1}{2}m_0 \omega _{z_1}^2(z+z_1)^2 & ,S_1(\rho ,z) \\
V_2= & \frac{1}{2}m_0 \omega _{\rho _2}^2 \rho ^2+
\frac{1}{2}m_0 \omega _{z_2}^2 (z-z_2)^2 & ,S_2(\rho ,z)
\end{array}
\end{equation} 
or

\begin{equation}
\omega _{\rho _1}^2 \rho ^2+\omega _{z_1}^2(z+z_1)^2=
\omega _{\rho _2}^2 \rho ^2+\omega _{z_2}^2(z-z_2)^2
\end{equation}
If one translates the $z-$coordinate:

\begin{equation}
z=z^{\prime }-k
\end{equation}
and separate the terms with respect to the powers of $\rho $ and
$z^{\prime }$, we get:

\begin{eqnarray}
(\omega _{\rho _1}^2-\omega _{\rho _2}^2)+
(\omega _{z_1}^2-\omega _{z_2}^2)z^{\prime 2}-
2k(\omega _{z_1}^2-\omega _{z_2}^2)z^{\prime}  
+(\omega _{z_1}^2-\omega _{z_2}^2)k^2 \nonumber \\
+2(\omega _{z_1}^2z_1+\omega _{z_2}^2z_2)z^{\prime}-
2(\omega _{z_1}^2z_1+\omega _{z_2}^2z_2)k+
\omega _{z_1}^2z_1^2-\omega _{z_2}^2z_2^2=0
\end{eqnarray}
We notice that:

\begin{equation}
z_1+z_2=R
\end{equation}
and after some simple calculations one obtains:

\begin{equation}
\frac{\rho ^2}{\displaystyle \frac{\omega _{z_1}^2 \omega _{z_2}^2R^2}
{(\omega _ {\rho _2}^2-\omega _{\rho _1}^2)(\omega _{z_2}^2-
\omega _ {z_1}^2)}} + 
\frac{\displaystyle (z+k)^2}{ \displaystyle
\frac{\omega _{z_1}^2 \omega _{z_2}^2R^2}
{(\omega _{z_2}^2-
\omega _ {z_1}^2)^2}} =1 
\end{equation}
which defines a so called unique center surface, here namely an ellipsoid
with semiaxes:

\begin{eqnarray}
 a_m & = &  \displaystyle \frac{\omega _{z_1} \omega _{z_2} R}
{\omega _{z_2}^2-\omega _{z_1}^2} \nonumber \\
b_m & = &  \displaystyle \frac{\omega _{z_1} \omega _{z_2}R}
{[(\omega _{\rho _2}^2-\omega _{\rho _1}^2)(\omega _{z_2}^2 -\omega _{z_1}^2)]
^{\frac{1}{2}}}
\end{eqnarray}

Since the origin of this surface is at $(\rho =0, z^{\prime}=0)$, its center
with respect to the arbitrary origin (against which the potentials are expressed)
is on the symmetry axis at:

\begin{equation}
z_{0m}=k= \frac{a_2^2z_1+a_1^2z_2}{a_1^2-a_2^2}
\end{equation}
What we have obtained are the coordinates $(z_{0m},0)$ and the semiaxes
$(a_m,b_m)$ of a matching potential ellipsoid (MPE). On its surface
$V_1(\rho ,z)=V_2(\rho ,z)$. The heavy deformed fragment $O_1$ is acting
through $V_1(\rho ,z)$ outside the MPE. The light fragment part which emerges
from the parent is contained inside MPE (the shaded region in Fig. 3). Hence
its action on the Hilbert space, $V_2(\rho ,z)$, goes all over the matching ellipsoid
volume.

The same demonstration is valid for $V_1(\rho ,z)=V_{g1}(\rho ,z)$, and
$V_2(\rho ,z)=V_{g2}(\rho ,z)$, as far as the neck potential shares its action
space with the fragments. First condition:

\begin{equation}
V_1(\rho ,z)=V_{g1}(\rho,z)
\end{equation}
or
\begin{equation}
\frac{1}{2}m_0\omega _{\rho _1}^2 \rho ^2+
\frac{1}{2}m_0\omega _{z_1}^2(z+z_1)^2=
2V_0-[\frac{1}{2}m_0\omega _g^2(\rho -\rho _3)^2+\frac{1}{2}m_0\omega _g^2
(z-z_3)^2]
\end{equation}
yields:

\begin{eqnarray}
(m_0\omega _{\rho _1}^2+m_0\omega _g^2)\rho ^2+(m_0\omega _{z_1}^2+
m_0\omega _g^2)z^2-2m_0\omega _g^2\rho _3 \rho + \nonumber \\
2(m_0 \omega _{z_1}^2z_1-m_0\omega _g^2z_3)z+
m_0\omega _{z_1}^2z_1^2+m_0\omega _g^2(\rho _3^2+z_3^2)-4V_0=0
\end{eqnarray}
This equation is of the type:

\begin{equation}
a_{11}\rho ^2+a_{22}z^2+2a_1\rho +2a_2z+a=0
\end{equation}
To describe a unique center surface, the normalized determinant of the
equation:

\begin{equation}
\frac{\Delta}{\delta}=a-\frac{a_1^2}{a_{11}}-\frac{a_2^2}{a_{22}} \nonumber
\end{equation}
must be positive, which is the casefor our situation. Then the center of the surface
has the coordinates:

\begin{eqnarray}
\rho _{0m1}= \frac{
\left| \begin{array}{cc}  a_{12} & a_1 \\ a_{22} & a_2 
\end{array} \right|}
{\left| \begin{array}{cc} a_{11} & a_{12} \\ a_{21} & a_{22}
\end{array} \right|    } & &
z_{0m1}=- \frac{
\left| \begin{array}{cc} a_{11} & a_1 \\ a_{21} & a_2
\end{array} \right|
}
{\left| \begin{array}{cc} a_{11} & a_{12} \\ a_{21} & a_{22}
\end{array} \right| }
\end{eqnarray} 
After some simple calculations one obtains:

\begin{eqnarray}
\rho _{0m1}=\frac{\omega _g^2 \rho _3}{\omega _{\rho _1}^2+\omega _g^2} & &
z_{0m1}=\frac{\omega _g^2z_3-\omega _{z_1}^2z_1}{\omega _{z_1}^2+\omega _g^2}
\end{eqnarray}
This surface is again an ellipsoid with semiaxes:

\begin{eqnarray}
a_{m1}=\left( -\frac{\Delta}{\delta} \frac{1}{a_{22}} \right)^{1/2} & &
b_{m1}=\left( -\frac{\Delta}{\delta} \frac{1}{a_{11}} \right)^{1/2}
\end{eqnarray}
and represents the matching potential ellipsoid (MPE1) between $V_1(\rho ,z)$ and
$V_{g1}(\rho ,z)$. On its surface, the two values equal each other. Obviously,
the tangent point between this first matching potential ellipsoid and the heavy
fragment shape is the same as the tangent point between the neck sphere and the
heavy fragment. In the interior of MPE1, the DTCSM potential is $V_{g1}(\rho ,z)$.
Following the same arguments for:

\begin{equation}
V_2(\rho ,z)=V_{g1}(\rho ,z)
\end{equation}
one finds:

\begin{eqnarray}
\rho _{0m2}=\frac{\omega _g^2 \rho _3}{\omega _{\rho _2}^2+\omega _g^2} & &
z_{0m2}=\frac{\omega _g^2z_3-\omega _{z_2}^2z_2}{\omega _{z_2}^2+\omega _g^2}
\end{eqnarray}
for the center of the second neck matching potential ellipsoid (MPE2), and the
same expressions with $V_2(\rho ,z)$ corresponding values instead of
$V_1$ hold for the semiaxes $a_{m2}$ and $b_{m2}$.

In the upper part of Fig. 3, the three matching ellipsoids - the large shaded
one for $V_1=V_2$ (MPE) and the two grey ones for $V_1=V_{g1}$ (MPE1) and
$V_2=V_{g1}$ (MPE2) respectively - are drawn. In this way DTCSM potential spans
the whole Hilbert space as follows: within MPE it is $V_2(\rho ,z)$ which is active;
within the grey area formed by the two neck potential matching ellipsoids,
MPE1 and MPE2, we have $V_{g1}(\rho ,z)$; in the interior of the nuclear shape,
between the neck surface and the fragment ellipsoids surfaces (tiny black region)
the potential is $V_{g2}(\rho ,z)$. The rest of the space is under $V_1(\rho ,z)$
action.

What is achieved in this section is spanning the Hilbert space by DTCSM potential,
passing continuously from the potential generated by the heavy fragment $V_1$ to
the lighter one $V_2$ and allowing also for the neck to be smoothly shaped by $V_g$.
There are no sharp cusps in the potential values. Since we work in axial symmetry,
the two neck matching ellipsoids MPE1 and MPE2 are in fact rolling around the symmetry
axis. Each of them shapes a thorus around the necking region, where $V_{g1}$ is
the DTCSM potential.

On the lower part of Fig. 3 geometrical points which will be used for further
calculations are marqued: $\rho _{m1s}(z)$ and $\rho _{m1j}(z)$ are MPE1 surface
functions, $\rho _{m2s}(z)$ and $\rho _{m2j}(z)$ hold for MPE2, and $\rho _m(z)$
is the surface of MPE. Other points of interest are: $z_{c1}$ and $z_{c2}$ - the
two tangent points of the neck {\it and} MPE1 and MPE2 with the fragment ellipsoids,
$z_{im}$ - the intersection of MPE with the symmetry axis, $z_{x1l}$ and $z_{x2r}$
- the left margin of MPE1 and the right margin of MPE2, $z_s$ - the crossing point
between the two fragment ellipsoids, $z_{mint}$ - upper crossing point between the
three matching ellipsoids

Fig. 4 shows how the three potentials work together.
 The values of $V_1(\rho ,z)$, $V_2(\rho ,z)$ and $V_{g1}(\rho ,z)$
are taken at $\rho =\rho _1(z_{c1})$ (upper part of the figure), and 
$\rho =\rho _2(z_{c2})$, hence at the tangent points of the neck sphere. Evolution
is shown as the distance $R$ (or $z$-variable) increases. The values are taken 
at a matching surface point, thus the potentials are tangent all the time. One can
observe how $V_1$ and $V_2$ wells are separating with increasing $R$. The matching
surfaces are composed of the whole set of tangent points at every $\rho $ distance
from the symmetry axis.

The way two of the potentials, $V_1(\rho ,z)$ and $V_{g1}(\rho ,z)$, behave along
the $O_1O_3$ line between their two centers is given in Fig. 5. The crossing points
of the curves correspond to $z_{c1}$ and its opposite along MPE1.
Inside the neck sphere, hence inside MPE1 too, $V_{g1}>V_1$.

\section{Diagonalization basis}

This section will browse succintly the main steps to be taken in order
to find a normalized functions set, since details about this issue have
already been published \cite{mar,bad}.

The new feature in the DTCSM eigenvalue problem is the shape-dependency
of the potentials: $V_{DTCSM}$, the spin-orbit term $V_{\mathbf \Omega 
\mathbf s}$ 
and the $V_{\mathbf \Omega ^2}$ term ($\mathbf \Omega$ is preferred here
as a notation since the angular momentum operator, as being shape dependent,
will be different from the usual $\mathbf l$). The total Hamiltonian:

\begin{equation}
H_{DTCSM}=-\frac{\hbar^2}{2m_0} \Delta +V_{DTCSM}(\rho ,z)+
V_{\mathbf \Omega \mathbf s }+V_{\mathbf \Omega ^2}
\end{equation}   
is obviously not separable. A basis is needed and diagonalization of oscillator
potential differences and of angular momentum dependent operators has to be
performed.

A separable Hamiltonian is obtained if one takes $\omega _{\rho _1} =
\omega _{\rho _2} =\omega _1$, with no $\mathbf l \mathbf s$ and $\mathbf l^2$
terms, hence a potential like:

\begin{equation} 
V^{(d)}(\rho ,z)= \left \{ \begin{array}{ccc}
V _1^{(d)}(\rho ,z)= & \frac{1}{2}m_0 \omega _1^2 \rho ^2
+\frac{1}{2}m_0 \omega _1^2(z+z_1)^2 & ,z \le 0 \\
V_2^{(d)}(\rho ,z)= & \frac{1}{2}m_0\omega _1^2 \rho ^2+
\frac{1}{2}m_0 \omega _2^2(z-z_2)^2 & ,z \ge 0
\end{array} \right.
\end{equation}
This is an appropiate two-center potential for a sphere ($z \le 0$ )
intersected with a vertical spheroid. The origin ($z=0$) is the intersection plane.
As a result of variable separation, three known differential equations are
obtained for harmonic functions, Laguerre polynomial and Hermite function dependent
solutions \cite{bad}.

Using continuity conditions for $z$-dependent functions and their derrivatives
at $z=0$, the normalization condition and equalizing the energy on the symmetry axis
$E_z=\hbar \omega _1(\nu _1+0.5)=\hbar \omega _2 (\nu _2+0.5)$, $z$-quantum numbers
and normalization constants for the $z$-dependent eigenfunctions are calculated.
The $ \phi $ and $\rho $-dependent functions are straitforward calculated from the
differential equations. The final result is:
\begin{eqnarray}
\Phi _m(\phi) & = & \frac{1}{\sqrt{2\pi}}\exp{(im \phi)} \nonumber \\
R_{n_{\rho}}^{|m|}(\rho) & = & \sqrt{\frac{2\Gamma (n_{\rho}+1)
\alpha _1^2}{\Gamma (n_{\rho}+|m|+1)}}
\exp{\left (-\frac{\alpha _1^2\rho ^2}{2}\right )}
(\alpha _1^2\rho ^2)^{\frac{|m|}{2}}
L_{n_{\rho}}^{|m|}(\alpha _1^2\rho ^2) \\
Z_{\nu}(z) & = & \left \{ \begin{array}{clc}
C_{\nu _1}\exp{\left [ -\frac{\alpha_1 ^2(z+z_1)^2}{2}\right ]}
H_{\nu _1}[-\alpha _1(z+z_1)] &,&z<0 \nonumber \\
C_{\nu _2}\exp{\left [ -\frac{\alpha _2^2(z-z_2)^2}{2}\right ]}
H_{\nu _2}[\alpha _2(z-z_2)] &,& z\geq 0
\end{array}\right .
\end{eqnarray}
We notice that:
where $\Gamma $ is the gamma function, $L_n^m$is the $m$-order Laguerre polynomial,
$C_1$ and $C_2$ the normalization constants, $\nu _1$, $\nu _2$ the quantum numbers
along the symmetry axis, and $H_{\nu }$ is the Hermite function. We now have a basis
for further calculations. The eigenvalues for the diagonalized Hamiltonian  with
the potential $V^{(d)}$ are the oscillator energy levels for sphere+spheroid system:
\begin{equation}
E_{osc}^{(d)}=\hbar \omega _1(2n_{\rho }+\mid m \mid +1)+
\hbar \omega _{z_1}(\nu _1 +0.5)
\end{equation}

At this point we have a useful basis for the calculation of $H_{DTCSM}$ matrix
elements.

\section{$H_{DTCSM}$ operators}

In order to obtain the DTCSM energy levels, the matrix of the non-diagonal elements
should be constructed. Then, adding $E_{osc}^{(d)}$ as diagonal terms, after 
diagonalization we obtain the eigenvalues.

\subsection{$V_{DTCSM}(\rho ,z)$ - terms}

This subsection is devoted to build the operators for the necked-in deformed
two-center oscillator levels calculation. We start wiht the operators leading to the
appropiate energies for $V_1(\rho ,z)$. The difference $V_1(\rho ,z)-V^{(d)}(\rho ,z)$
on the volume $v_1$ and added to $E^{(d)}(n_{\rho },\nu ,m)$ produce the suitable
matrix elements for $V_1$:
\begin{eqnarray}
\Delta V_1 & = & \Delta V_1(- \infty ,z_{im})+\Delta V_1(z_{im},0)+
\Delta V_1(0,z_{0m}+a_m)+ \Delta V_1(z_{0m}+a_m, \infty )   \\
 & & -\Delta V_1(z_{x1l} \div z_{c1}; \rho _{m1j} \div \rho _{m1s})-
\Delta V_1(z_{c1} \div z_s ; \rho _1 \div \rho _{m1s})
-\Delta V_1(z_s \div z_{mint} ; \rho _m \div \rho _{m1s}) \nonumber
\end{eqnarray}
The positive terms account for the influence of $V_1(\rho ,z)$ as if $V_1$
would act alone everywhere but MPE. The negative terms are under $V_g(\rho ,z)$
action. In parenthesis are the geometrical limits for every term. Note that for
those which have also $\rho $-dependent limits, $\rho = \rho (z)$. The limits
can be easily tracked in Fig. 3. What is left after the subtraction is the difference
from $V_1(\rho ,z)$ to $V^{(d)}(\rho ,z)$ within the $v_1$ volume. To ease calculations,
one uses as a notation:
\begin{equation}
\Delta V_1^{(p)}  =  \Delta V_1(- \infty ,z_{im})+\Delta V_1(z_{im},0)+
\Delta V_1(0,z_{0m}+a_m)+ \Delta V_1(z_{0m}+a_m, \infty )   
\end{equation}
These pozitive potential differences are now written in terms of the Heavyside 
step function; this will result in determining the exact limits where $V_1(
\rho ,z)$ is acting and, consequently, where it decides the oscillator
energy values :
\begin{eqnarray}
\Delta V_1^{(p)} & = & \frac{1}{2}(m_0 \omega _{\rho _1}^2-
m_0 \omega _1^2) \{ [ 1-\theta (z-z_{im}) ] +
\theta [z-(z_{0m}+a_m) ] \} \rho ^2 \nonumber \\
 & & +\frac{1}{2} \{ (m_0 \omega _{z_1}^2-m_0 \omega _1^2)(z+z_1)^2
[1-\theta (z-z_{im})] \nonumber \\
 & & +
[m_0 \omega _{z_1}^2(z+z_1)^2-m_0\omega _2^2(z-z_2)^2]
\theta [z-(z_{0m}+a_m)] \} \delta _{n_{\rho ^{\prime}}n_{\rho }} \\
 & & + \frac{1}{2}(m_0\omega _{\rho _1}^2-m_0 \omega _1^2)
\{ [ \theta (z-z_{im})-\theta (z) ] \nonumber \\
 & & +
[ \theta (z)-\theta [z-(z_{0m}+a_m) ] ] \} \rho ^2 \theta (\rho -\rho _m(z)) 
\nonumber \\
 & & + \frac{1}{2} \{(m_0 \omega _{z_1}^2 -m_0 \omega _1^2)(z+z_1)^2
[\theta (z-z_{im})-\theta (z) ]  \nonumber \\
 & & + [m_0 \omega _{z_1}^2(z+z_1)^2
-m_0\omega _2^2(z-z_2)^2 ][\theta (z)-\theta [z-(z_{0m}+a_m)] ] \}
\theta (\rho -\rho _m (z)) \nonumber
\label{eq:dv1}
\end{eqnarray} 
For the negative differences we have:

\begin{eqnarray}
\lefteqn{
\Delta V_1(z_i \div z_f; \rho _i(z) \div \rho _f(z))  = } \nonumber \\
 & &\frac{1}{2}m_0\omega _{\rho _1}^2 [\theta (z-z_i) \theta (z-z_f)]
\rho ^2 [\theta (\rho -\rho _i(z))-\theta [\rho -\rho _f (z))]  \\
 & & +\frac{1}{2}m_0 \omega _{z_1}^2(z+z_1)^2[\theta (z-z_i)-\theta (z-z_f)]
[\theta (\rho -\rho _i(z))-\theta (\rho - \rho _f(z))] \nonumber
\end{eqnarray}
where $z_i$, $z_f$, $\rho _i(z)$ and $\rho _f(z)$ take the corresponding
values and limit functions from Eq. \ref{eq:dv1}.

To obtain the terms where the light emerged fragment potential $V_2(\rho ,z)$
acts, one must subtract $V^{(d)}(\rho ,z)$ from $V_2(\rho ,z)$ within the $v_2$ 
volume. The result is:

\begin{eqnarray}
\lefteqn{
\Delta V_2 = } \nonumber \\
& & \Delta V_2(z_{im} \div 0; 0 \div \rho _m(z))
+\Delta V_2(0 \div z_{0m}+a_m; 0 \div \rho _m(z)) \nonumber \\
 & & - \Delta V_2(z_s \div z_{c2}; \rho _g(z) \div \rho _m(z))
-\Delta V_2(z_{mint} \div z_{x2r}; \rho _{m2j}(z) \div \rho _{m2s}(z))
\end{eqnarray}

Let us denote:

\begin{equation}
\Delta V_2^{(p)}=\Delta V_2 (z_{im} \div 0;0 \div \rho _m(z))+
\Delta V_2(0 \div z_{0m}+a_m;0 \div \rho _m(z))
\end{equation}

Then we have:

\begin{eqnarray}
\lefteqn{\Delta V_2^{(p)}=} \nonumber \\
 & & \frac{1}{2}(m_0 \omega _{\rho _2}^2-m_0 \omega _1^2) \{ \theta
[z-(z_{0m}+a_m)] \} \rho ^2 [1-\theta (\rho -\rho _m (z)) ] \nonumber \\
 & & +\frac{1}{2} \{ [ m_0 \omega _{z_2}^2(z-z_2)^2
-m_0 \omega _1^2(z+z_1)^2 ] [\theta (z-z_{im}) -\theta (z) ] \nonumber \\
 & & +(m_0 \omega _{z_2}^2-m_0 \omega _2^2)(z-z_2)^2 
\{ \theta (z) -\theta [z-(z_{0m}+a_m)] \} \}  
[1- \theta (\rho -\rho _m (z)) ]
\end{eqnarray} 
for the positive part; for the negative necking region which has to be subtracted
from $v_2$, the expression for any $\Delta V_2 (z_i \div z_f ; \rho _i(z)
\div \rho _f(z))$ is:

\begin{eqnarray}
\lefteqn { \Delta V_2 (z_i \div z_f; \rho _i \div \rho _f(z))=} \\
 & & \frac{1}{2}m_0 \omega _{\rho _2}^2 [\theta (z-z_i)-\theta (z-z_f)]
\rho ^2 [\theta (\rho - \rho _i(z)- \theta (\rho - \rho _f(z))] \nonumber \\
 & & + \frac{1}{2}m_0\omega _{z_2}^2(z-z_2)^2[\theta (z-z_i) - \theta (z-z_f)]
[\theta (\rho - \rho _i(z))-\theta (\rho -\rho _f(z)) ]
\end{eqnarray}
using the same convention for $z_i$, $z_f$, $\rho _i $ and $\rho _f(z)$ as for
$\Delta V_1$.

At this point, the neck potential $V_g(\rho ,z)$ is filling its volume $v_g$ without
any subtraction:

\begin{equation}
\Delta V_g=V_g
\end{equation}  

With the same notation as in Eq. \ref{eq:os} we have for the neck operators:

\begin{equation}
V_g=V_{g1}(v_{g1})+V_{g2}(v_{g2})
\end{equation}
Now every of them will be taken separately:

\begin{equation}
V_{g1}=V_{g1}(z_{x1l} \div z_{c1};\rho _ {m1j} \div \rho _{m1s})
+V_{g1}(z_{c1} \div z_{mint};\rho _g \div \rho _{m1s})
+V_{g1}(z_{mint} \div z_{x2r}; \rho _{m2j} \div \rho _{m2s})
\end{equation} 

\begin{equation}
V_{g2}=V_{g2}(z_{c1} \div z_s;\rho _1(z) \div \rho _g(z))
+V_{g2}(z_s \div z_{c2}; \rho _2(z) \div \rho _g(z))
\end{equation}
where again:

\begin{eqnarray}
\lefteqn{V_{g1}(z_i \div z_f;\rho _i \div \rho _f)=}  \nonumber \\
 & & 2V_0 [ \theta (z-z_i)- \theta (z-z_f) ]
[\theta (\rho - \rho _i(z))-\theta (\rho - \rho _f (z)) ] \nonumber \\
 & & -\frac{1}{2}m_0 \omega _g^2 [\theta (z-z_i)-\theta (z-z_f) ]
(\rho - \rho _3)^2[\theta (\rho - \rho _i(z))-\theta (\rho -\rho _f(z))]
\nonumber \\
 & & -\frac{1}{2}m_0\omega _g^2(z-z_3)^2[\theta (z-z_i)-\theta(z-z_f)]
[\theta (\rho -\rho _i (z))-\theta (\rho -\rho _f(z))]
\end{eqnarray}
and

\begin{equation}
V_{g2}(z_i \div z_f; \rho _i \div \rho _f)=
V_0 \{[\theta (z-z_i)-\theta (z-z_f)]
[\theta (\rho -\rho _i(z))-\theta (\rho -\rho _f(z))] \}
\end{equation}
The final analytical expressions for the matrix elements corresponding
to the oscillator potential of DTCSM are given in Appendix A.

\subsection{Spin-orbit $\mathbf l \mathbf s $ and $\mathbf l^2 $ operators}

Special care will be devoted to the treatment of $V_{\mathbf l \mathbf s } $
and $V_{\mathbf l^2}$ matrix elements. Because of the dependence of the 
angular momentum term on different space regions ( $\sim \nabla V \times \mathbf p $),
hence on different mass regions when asymmetry $A_1/A_2$ comes in, the 
anticommutator is used to assure hermicity for the operators:

\begin{equation} 
V_{\mathbf l \mathbf s}= \left \{ \begin{array}{cc}
-\left \{\displaystyle \frac{\hbar}{m_0 \omega _{01}} \kappa _1(\rho ,z),
(\nabla V \times \mathbf p )\mathbf s \right \} & ,A_1-region \nonumber \\
-\left \{\displaystyle \frac{\hbar }{m_0 \omega _{02}} \kappa _2(\rho ,z),
(\nabla V \times \mathbf p ) \mathbf s \right \} & , A_2-region
\end{array} \right.
\label{eq:ls}
\end{equation}
and

\begin{equation} 
V_{\mathbf l^2}= \left \{ \begin{array}{cc}
-\left \{\displaystyle \frac{\hbar}{m_0 ^2 \omega _{01}^3}
 \kappa _1 \mu _1(\rho ,z),
(\nabla V \times \mathbf p )^2 \right \} & ,A_1-region \nonumber \\
-\left \{\displaystyle \frac{\hbar }{m_0^2 \omega _{02}^3} 
\kappa _2 \mu _2(\rho ,z),
(\nabla V \times \mathbf p )^2  \right \} & , A_2-region
\end{array} \right.
\end{equation}

Basically the same treatment as for oscillator terms is valid in this case.
The $\kappa _1(\rho ,z)$ and $\mu _1(\rho ,z)$ are the strength function
parameters for the $V_1(\rho ,z)$ region, whereas $\kappa _2(\rho ,z)$ and
$\mu _2 (\rho, z)$ are active for the $V_2(\rho ,z)$ one.

The new feature here is the use of the anticommutator for the operators 
containing Heavyside function combinations, to confine the action to $v_1$,
$v_2$ and $v_g$; these function combinations are exactly the ones which have been
used for $\Delta V_1$, $\Delta V_2$ and $\Delta V_g$ terms.

Since the creation and anihilation angular momentum operators become shape-frequency
dependent, notations will be changed as follows:

\begin{eqnarray}
\mathbf l^+ \rightarrow \mathbf \Omega ^+ &
\mathbf l^- \rightarrow \mathbf \Omega ^- &
\mathbf l_z \rightarrow \mathbf \Omega _z \nonumber 
\end{eqnarray}
so that

\begin{eqnarray}
\mathbf l \mathbf s & \rightarrow & 
\frac{1}{2}(\mathbf \Omega ^+ \mathbf s^- +
\mathbf \Omega ^- \mathbf s^+) +\mathbf \Omega _z \mathbf s_z
\end{eqnarray}
Then for the potentials of the spin-orbit term to be diagonalized one reads:

\begin{equation}
V_{\mathbf \Omega \mathbf s}=
V_{\mathbf \Omega \mathbf s}(v_1)+
V_{\mathbf \Omega \mathbf s}(v_2)+
V_{\mathbf \Omega \mathbf s}(v_g)
\end{equation}
where:

\begin{equation}
V_{\mathbf \Omega \mathbf s}(v_1)=
-\frac{\hbar}{m_0 \omega _{01}}
\kappa _1 \{\mathbf \Omega \mathbf s,(v_1) \}
\end{equation}

\begin{equation}
V_{\mathbf \Omega \mathbf s}(v_2)=
-\frac{\hbar}{m_0 \omega _{02}}
\kappa _2 \{\mathbf \Omega \mathbf s,(v_2) \}
\end{equation}

\begin{equation}
V_{\mathbf \Omega \mathbf s}(v_g)=
-\frac{\hbar}{m_0 \omega _{01}}
\kappa _1 \{\mathbf \Omega \mathbf s,(v_g^{(1)}) \}
-\frac{\hbar}{m_0 \omega _{02}}
\kappa _2 \{\mathbf \Omega \mathbf s,(v_g)^{(2)} \}
\end{equation}
where $v_g^{(1)}$ and $v_g^{(2)}$ are the neck matching ellipsoids
volumes on $A_1$ and $A_2$ side, respectively.

On the other hand, the operators depend on the region they exert themselves
through the potentials, according to Eq. \ref{eq:ls}. Thus we have:

\begin{eqnarray}
\mathbf \Omega ^+(v_1) & = &
- e^{i \varphi}
\left [ \frac{\partial V_1(\rho ,z)}{\partial \rho}
        \frac{\partial }{\partial z} 
       -\frac{\partial V_1(\rho ,z)}{\partial z}
        \frac{\partial }{\partial \rho }
       -\frac{i}{\rho }\frac{\partial V_1(\rho ,z)}{\partial z}
                       \frac{\partial }{\partial \varphi} \right ] \nonumber \\
 & = & - e^{i \varphi}  \left [
m_0 \omega _{\rho _1}^2 \rho  \frac{\partial }{\partial z}
-m_0 \omega _{z_1}^2(z+z_1)\frac{\partial }{\partial \rho }
-\frac{i}{\rho }m_0\omega _{z_1}^2(z+z_1)\frac{\partial }{\partial \varphi} \right ]
\end{eqnarray}
 
\begin{eqnarray}
\mathbf \Omega ^-(v_1) & = & 
e^{-i \varphi}
\left [ \frac{\partial V_1(\rho ,z)}{\partial \rho}
        \frac{\partial }{\partial z} 
       -\frac{\partial V_1(\rho ,z)}{\partial z}
        \frac{\partial }{\partial \rho }
       +\frac{i}{\rho }\frac{\partial V_1(\rho ,z)}{\partial z}
                       \frac{\partial }{\partial \varphi} \right ] \nonumber \\
 & = &  e^{-i \varphi}  \left [
m_0 \omega _{\rho _1}^2 \rho  \frac{\partial }{\partial z}
-m_0 \omega _{z_1}^2(z+z_1)\frac{\partial }{\partial \rho }
+\frac{i}{\rho }m_0\omega _{z_1}^2(z+z_1)\frac{\partial }{\partial \varphi} \right ]
\end{eqnarray}

\begin{eqnarray}
\mathbf \Omega _z(v_1) & = & 
  -\frac{i}{\rho }\frac{\partial V_1}{\partial \rho }\frac{\partial }{\partial 
\varphi } \nonumber \\
 & = & -im_0\omega _{\rho _1}^2\frac{\partial }{\partial \varphi }
\end{eqnarray}
with

\begin{equation}
\mathbf \Omega \mathbf s (v_1)=
\frac{1}{2}(\mathbf \Omega ^+(v_1)\mathbf s^- 
+\mathbf \Omega ^- \mathbf s^+)+\mathbf \Omega _z(v_1)
\mathbf s_z
\end{equation}
The same aplies for $(v_2)$ and $(v_g)$ spin-orbit operators terms, and the expressions
are:

\begin{eqnarray}
\mathbf \Omega ^+(v_2) & = &
  - e^{i \varphi}  \left [
m_0 \omega _{\rho _2}^2 \rho  \frac{\partial }{\partial z}
-m_0 \omega _{z_2}^2(z-z_2)\frac{\partial }{\partial \rho }
-\frac{i}{\rho }m_0\omega _{z_2}^2(z-z_2)\frac{\partial }
{\partial \varphi} \right ] \nonumber \\
\mathbf \Omega ^-(v_2) & = & 
   e^{-i \varphi}  \left [
m_0 \omega _{\rho _2}^2 \rho  \frac{\partial }{\partial z}
-m_0 \omega _{z_2}^2(z-z_2)\frac{\partial }{\partial \rho }
+\frac{i}{\rho }m_0\omega _{z_2}^2(z-z_2)\frac{\partial }
{\partial \varphi} \right ] \nonumber \\
\mathbf \Omega _z(v_2) & = & 
  -im_0\omega _{\rho _2}^2\frac{\partial }{\partial \varphi }
\end{eqnarray}
and:

\begin{eqnarray}
\mathbf \Omega ^+(v_{g1}) & = &
  - e^{i \varphi}  \left [
m_0 \omega _g^2 (\rho -\rho _3)  \frac{\partial }{\partial z}
-m_0 \omega _g^2(z-z_3)\frac{\partial }{\partial \rho }
-\frac{i}{\rho }m_0\omega _g^2(z-z_3)\frac{\partial }
{\partial \varphi} \right ] \nonumber \\
\mathbf \Omega ^-(v_{g1}) & = & 
   e^{-i \varphi}  \left [
m_0 \omega _g^2 (\rho -\rho _3)  \frac{\partial }{\partial z}
-m_0 \omega _g^2(z-z_3)\frac{\partial }{\partial \rho }
+\frac{i}{\rho }m_0\omega _g^2(z-z_3)\frac{\partial }
{\partial \varphi} \right ] \nonumber \\
\mathbf \Omega _z(v_{g1}) & = & 
  im_0\omega _g^2\frac{\rho -\rho _3}{\rho } 
\frac{\partial }{\partial \varphi }
\end{eqnarray}
and $\mathbf \Omega \mathbf s(v_{g2})=0$, since $V_g(v_{g2})=V_0=cst$.
With the help of these operators, what is to be calculated now for the
spin-orbit shape-dependent matrix elements reads, for $(v_1)$:

\begin{equation}
V_{\mathbf \Omega \mathbf s}(v_1)=
-\frac{\hbar }{m_0 \omega _{01}} \kappa _1 
\{ \mathbf \Omega \mathbf s(v_1), (v_1) \}
\end{equation}
where:

\begin{eqnarray}
\lefteqn{ \{ \mathbf \Omega \mathbf s(v_1),(v_1) \} =} \nonumber \\
 & & \{\mathbf \Omega \mathbf s(v_1),[1-\theta (z-z_{im})] \}
+ \{ \mathbf \Omega \mathbf s(v_1), \theta [z-(z_{0m}+a_m)] \} 
\nonumber \\
 & & + \{ \mathbf \Omega \mathbf s(v_1),[\theta (z-z_{im})-
\theta [z-(z_{0m}+a_m)]]\theta (\rho -\rho _m(z)) \} \nonumber \\
 & & -\{ \mathbf \Omega \mathbf s(v_1),[\theta (z-z_{x1l})-
\theta (z-z_{c1})][\theta (\rho -\rho _{m1j}(z))-
\theta (\rho -\rho _{m1s}(z))] \} \nonumber \\
 & & - \{ \mathbf \Omega \mathbf s(v_1),[\theta (z-z_{c1})
-\theta (z-z_s)][\theta (\rho -\rho _1(z))-
\theta (\rho -\rho _{m1s}(z))] \} \nonumber \\
 & & - \{ \mathbf \Omega \mathbf s(v_1),
[\theta (z-z_s)-\theta (z-z_{mint})]
[\theta (\rho -\rho _m(z))-\theta (\rho -\rho _{m1s}(z)] \}
\end{eqnarray}
For the $V_2(\rho ,z) $ controlled region we have:

\begin{equation}
V_{\mathbf \Omega \mathbf s}(v_2)=
-\frac{\hbar }{m_0 \omega _{02}} \kappa _2
\{ \mathbf \Omega \mathbf s(v_2),(v_2) \}
\end{equation}
where:

\begin{eqnarray}
\lefteqn{ \{ \mathbf \Omega \mathbf s(v_2),(v_2) \} = } \nonumber \\
 & & \{ \mathbf \Omega \mathbf s(v_2),[\theta (z-z_{im})-
\theta [z-(z_{0m}+a_m)]][1-\theta (\rho -\rho _m(z))] \} \nonumber \\
 & & -\{ \mathbf \Omega \mathbf s(v_2),[\theta (z-z_s)-\theta (z-z_{c2})]
[\theta (\rho -\rho _2(z))-\theta (\rho -\rho _g(z))] \} \nonumber \\
 & & - \{ \mathbf \Omega \mathbf s(v_2),[\theta (z-z_{gm})
-\theta (z-z_{mint})][\theta (\rho -\rho _g(z))-\theta (\rho -\rho _m(z))] \}
\nonumber \\
 & & -\{ \mathbf \Omega \mathbf s(v_2),[\theta (z-z_{mint})-\theta (z-z_{x2r})]
[\theta (\rho -\rho _{m2j}(z))-\theta (\rho -\rho _{m2s}(z))] \}
\end{eqnarray}

Finally, the neck potential dependent spin-orbit interaction looks like:

\begin{equation}
V_{\mathbf \Omega \mathbf s}(v_g)=
-\frac{\hbar}{m_0 \omega _{01}}\kappa _1\{\mathbf \Omega \mathbf s(v_{g1}),
(v_{g1}^{(1)}) \}-\frac{\hbar }{m_0 \omega _{02}}\kappa _2 \{\mathbf \Omega
\mathbf s(v_{g1}),(v_{g1}^{(2)}) \}
\end{equation}
where: 

\begin{eqnarray}
\lefteqn{\{ \mathbf \Omega \mathbf s (v_{g1}),(v_{g1}^{(1)}) \} = } \nonumber \\
 & &\{ \mathbf \Omega \mathbf s (v_{g1}),[\theta (z-z_{x1l})-\theta (z-z_{c1})]
[\theta (\rho -\rho _{m1j}(z))-\theta (\rho -\rho _{m1s}(z))] \} \nonumber \\
 & & +\{ \mathbf \Omega \mathbf s(v_{g1}),
[\theta (z-z_{c1})-\theta (z-z_{mint})][\theta (\rho -\rho _g(z))-
\theta (\rho - \rho _{m1s}(z))] \}
\end{eqnarray}    

\begin{eqnarray}
\lefteqn{ \{ \mathbf \Omega \mathbf s(v_{g1}),(v_{g1}^{(2)}) \}= } \nonumber \\
 & & \{ \mathbf \Omega \mathbf s (v_{g1}),
[\theta (z-z_{mint})-\theta (z-z_{x2r})]
[\theta (\rho -\rho _{m2j}(z))-\theta (\rho -\rho _{m2s}(z))] \}
\end{eqnarray}         

For the $\mathbf l^2$ $\rightarrow $ $\mathbf \Omega ^2 $ term the usual expression
works:

\begin{equation}
\mathbf \Omega ^2 = \frac{1}{2}(
\mathbf \Omega ^+ \mathbf \Omega ^- + \mathbf \Omega ^-
\mathbf \Omega ^+)+\mathbf \Omega _z^2
\end{equation} 
and the operator for the whole space is:

\begin{eqnarray}
V_{\mathbf \Omega ^2}=
-\frac{1}{2} \left \{ \frac{\hbar}{m_0^2 \omega _{0i}^3}
\kappa _i \mu _i (\rho ,z), \mathbf \Omega ^2 \right \} & & ,i=1,2
\end{eqnarray}

Strength coefficients $\kappa _1 \mu _1 (\rho ,z)$ and
$\kappa _2 \mu _2 (\rho, z) $ are the same $(\rho ,z)$ - dependent functions
as $\kappa _1(\rho ,z)$ and $\kappa _2(\rho ,z)$ in the spin-orbit term. Making use
of the unity operator $I= \sum_{''}|'' \rangle \langle '' | $, many expressions can be
derived to obtain an easy workable decomposition of the lengthy matrix elements of
$V_{\mathbf \Omega ^2}$. The chosen one follows, in order to make use of the
already calculated expression for $\mathbf \Omega \mathbf s $ term:

\begin{eqnarray}
\lefteqn{\langle '| \{ f(\rho ,z),\mathbf \Omega ^+ \mathbf \Omega ^- \} |
\rangle = } \nonumber \\
 & &\sum_{''} [2
\langle ' | \{ f(\rho ,z),\mathbf \Omega ^+ \} | '' \rangle+
\langle ' |(\mathbf \Omega  ^+f(\rho ,z)) |'' \rangle ]
\langle '' |\mathbf \Omega ^- | \rangle + 
\langle ' | \mathbf \Omega ^+ | '' \rangle
\langle '' | ( \mathbf \Omega ^-f(\rho ,z))  | \rangle
\end{eqnarray}

\begin{eqnarray}
\lefteqn{\langle '| \{ f(\rho ,z),\mathbf \Omega ^- \mathbf \Omega ^+ \} |
\rangle = } \nonumber \\
 & &\sum_{''} [2
\langle ' | \{ f(\rho ,z),\mathbf \Omega ^- \} | '' \rangle+
\langle ' |(\mathbf \Omega  ^-f(\rho ,z)) |'' \rangle ]
\langle '' |\mathbf \Omega ^+ | \rangle + 
\langle ' | \mathbf \Omega ^- | '' \rangle
\langle '' | ( \mathbf \Omega ^+f(\rho ,z))  | \rangle
\end{eqnarray}

The needed expressions for the $\mathbf \Omega \mathbf s $ and $\mathbf
\Omega ^2$ matrix elements are detailed in Appendix B.

Now the total matrix elements for DTCSM can be calculated as:

\begin{equation}
 \langle i | DTCSM | j \rangle = 
  E_{osc}^{(d)}(n_{\rho },|m|, \mu )+
\langle i | \Delta V_1 | j \rangle + \langle i | \Delta V_2 | j
\rangle  
  + \langle i | V_g | j \rangle + \langle i | V_{\mathbf \Omega
\mathbf s } | j \rangle + \langle i | V_{ \mathbf \Omega ^2} | j \rangle
\end{equation}

\section{Level schemes and shell effects}

A ten shell scheme matrix elements have been calculated, which means
a 440 $\times $ 440 matrix. After proper diagonalization, the levels
are obtained.

First, DTCSM spectra are computed for the superheavy fission reaction channel
$^{306}122 \rightarrow ^{198}W+^{108}Cd $, with the nuclei deformations
$\chi _{122}=0.9 $, $\chi _W=1. $ and $\chi _{Cd}=0.83 $. Semiaxes $a_i$ and
$b_i$ are calculated from corresponding deformation parameter $\beta _2$
for every fragment, and using the volume conservation condition 
($\beta _2^{^{198}W}=$0., $\beta _2^{^{108}Cd}=$0.135 \cite{mol}).
The reduced distance between centers is chosen to represent the elongation
parameter of the shape : $(R-R_i)/(R_f-R_i)$, where $R_i$ is the distance between
centers when the light emerging fragment is completely embeded in the parent nucleus,
and $R_f=a_1+a_2+2R_3$ represents the final distance between centers, when the neck
sphere is aligned with the fragments. 

A direct consequence of the variation of the neck parameter $R_3$ on
microscopic behaviour of a fission process is depicted in Fig. 6. Here the
shell corections are drawn for the five $R_3$ - parameter values, as a
function of the distance between centers. They are calculated with the
Strutinsky method. A shallow minimum at $R
\approx $ 2.5 fm suggest a small deformation of the $^{306}122$ ground
state. First bump around $R \approx $ 5 - 6 fm shows up in every case.
Second bump is also there, but its position changes in $R$-value and in
height (between $R=$12 fm and $R=$20 fm). The rather deep minimum ($ \approx
$ -5 MeV) for large neck radii (apparent on the
$R_3=$ 10 fm curve) is probably not manifested in $^{306}122$ fission, since
at this distance between centers, the system is out of the fission barrier
due to strong Coulombian repulsion.

Second example takes into account the mass asymmetry parameter. Again is the fission
of the superheavy $^{306}122$ through two channels this time:

\begin{eqnarray}
^{306}122  & \rightarrow & ^{198}W+^{108}Cd \nonumber \\
^{306}122 & \rightarrow & ^{154}Gd+^{152}Ce \nonumber 
\end{eqnarray}
The two corresponding level schemes are on the upper side of Fig. 7 
($\chi _{Gd} $ =0.72 for $\beta _2^{Gd}=$ 0.243 and $\chi _{Ce}=$ 0.71 for
$\beta _2^{Ce}=$0.261). They have been chosen because of the mass symmetry of the
second reaction against an asymmetric one. Calculations in this case preserved
$R_3=$4 fm. Shells are more clearly visible at $(R-R_i)/(R_f-R_i)$=1 in the symmetric
case, since asymptotically the levels are practically coincident for the two
quasi-equal fragments. The lower part of the figure represents the shell corrections
calculated with the above level schemes as an input. The first bump is shifted
towards smaller $R$ for symmetry. A second minimum as deep as almost the ground
state along the symmetric channel is an indication of a possible shape isomer.
Second maximum appears only for the asymmetric reaction. Asymptotically there is a
$\Delta E \approx $4 MeV difference in between the two channels, suggesting a possible
fission path along the mass asymmetry degree of freedom.

The last studied case is the fission of $^{252}Cf$ ($\chi _{Cf}=$0.73 for
$\beta _2^{Cf}=$0.236) through two favorized channels in Fig. 8:

\begin{eqnarray}
^{252}Cf  & \rightarrow & ^{160}Sm+^{92}Kr \nonumber \\
^{252}Cf & \rightarrow & ^{144}Ba+^{108}Mo \nonumber 
\end{eqnarray}
These two reactions have been chosen because they lie within the maxima
of the fission fragment mass distribution of $^{252}Cf$ \cite{wah}.
The fragment deformation parameters considered in calculations are
$\chi _{Sm}$=0.68 for $\beta _2^{Sm}$=0.290;
$\chi _{Kr}$=0.73 for $\beta _2^{Kr}$=0.228;
$\chi _{Ba}$=0.8 for $\beta _2^{Ba}$=0.164;
$\chi _{Mo}$=0.65 for $\beta _2^{Mo}$=0.333.
Level schemes are rather likely in the first part of the splitting, a fact that can 
be observed from the small diferences in the shell corrections up to $R \approx $7 fm.
Beyond this distance between centers, the effect of the wells separation
becomes more and more important, fragment individuality imposing the difference.
If only for the shell corrections, $^{160}Sm+^{92}Kr$ seem to be a more probable
fission channel, but it beyond the purpose of this work to do a fission process
analysis. Separation occurs around 18 fm, for approximately both of the reactions.
As it is already known, lower energy shells of the fragments are the first to unveil
asymptotically, whereas the higher states of the two systems still interact. This is
an expected result since the potentials separate starting with the bottom of the wells,
while the upper part of the potentials still overlap.   

\section{Conclusions}

The deformed two-center shell model presented in this work describes the evolution
of single particle levels from one parent potential well to the two fragments ones.

The introduction of fragment ellipsoidal deformation in the two-center oscillator
wells and further on, in the spin-orbit and $\mathbf l^2$ operators enables a more
realistic calculation of the two interacting quantum systems. This new form of
spin-orbit $\mathbf \Omega \mathbf s$ and $\mathbf \Omega ^2$ allows for the
angular momentum dependent operators to follow the exact sequence of shapes
throughout the splitting process. In such a way ellipsoidal degrees of freedom are
considered within spin-orbit interaction.

The new necking-in dependent microscopic potential results in considering the neck
degree of freedom into the level scheme calculation. As has been shown, the last part
of the fision process is influenced by the difference in neck radii. This fact has
important consequences on the potential barriers and fission paths on a potential
energy surface calculated within more deformation degrees of freedom, necking included.

A new way of treating two partially overlapping nuclei has been introduced by making
use of matching potential surfaces. It is only in this way that {\it continuity}
between different regions of potential influence is assured. The potential pass
smoothly from $V_1$ to $V_g$, to $V_2$ and so on without any cusp in its value and
without the introduction of arbitrary geometric transition functions.
For zero neck radius one obtains the fusion-like type of shapes.

All these facts make the presented model suitable for the study of fission channels
and cluster decay phenomena and calculation of potential energy surfaces in fission. 

Also the model provides the analysis of possible deformed target-projectile doors
toward fusion processes in heavy ion reactions. In nucleus-nucleus collisions,
calculations of two fusioning deformed partners yield the single particle  spectra
for the important (yet avoided up to now) region of overlapping shapes. 

\appendix
\section{DTCSM oscillator matrix elements}

Formulas for the two deformed oscillators with necking-in are presented to
enable $\langle '|\Delta V_1 | \rangle $, $\langle ' | \Delta V_2 | \rangle $
and $ \langle ' | \Delta V_g | \rangle $ matrix elements to be calculated.
The spin-dependent part is not included, since it introduce only a $\delta _{s's}$
factor.

We preserve the same notations as in section 6. $\langle i | j \rangle 
_{\alpha }^{\beta }$
 stands for
the integral calculated from $\alpha $ to $ \beta $ limits; $| n_z \rangle $
is the $Z_{\nu _{n_z}}(z) $ wave function; $| n_{\rho } m n_z \rangle $
is the total $\Phi _m R_{n_{\rho }}^{|m|}Z_{\nu _{n_z}} $ wave function of the basis.

The following notations will be available from now on:
\begin{equation}
I(n_{\rho}',n_{\rho },m',m,m_g',m_g,m_p,t)=
\left [\frac{\Gamma (n_{\rho }'+1) \Gamma (n_{\rho }+1)}
{\Gamma (n_{\rho }'+m_g') \Gamma (n_{\rho }+m_g)} \right ]^
{\frac {1}{2}}\int_t^{\infty } e^{-x}x^{m_p}L_{n_{\rho }'}^{|m'|}(x)
L_{n_{\rho }}^{|m|}(x)dx
\end{equation}
which is computed by Gauss-Laguerre nummerical integration procedure, and:

\begin{equation}
G(n_{\rho }',n_{\rho },m',m,t)=
\left [ \frac{\Gamma (n_{\rho }'+1)\Gamma (n_{\rho }+1)}
{\Gamma (n_{\rho }'+|m'|+1)\Gamma(n_{\rho}+|m|+1)} \right ]^{\frac{1}{2}}
2\alpha _1 t^{m+m'+\frac{1}{2}}e^{-t}L_{n_{\rho }'}^{|m'|}(t)
L_{n_{\rho }}^{|m|}(t)
\end{equation}
where:

\begin{eqnarray}
\alpha _{\rho _i}=\left ( \frac{m_0 \omega _{\rho _i}}{\hbar} \right )
^{\frac{1}{2}} &
\alpha _{z_i}=\left ( \displaystyle \frac{m_0 \omega _{z_i}}{\hbar} \right )
^{\frac{1}{2}} &
\alpha _1=\left (\frac{m_0 \omega _1}{\hbar } \right )^{\frac{1}{2}}
\end{eqnarray}
where $\omega _1 $ is the corresponding frequency for the sphere with the same
volume. Relations between ellipsoids frequencies and the one for the sphere can
easily be obtained from volume conservation $ab^2=R_i^3$, when $a$, $b$ are the 
semiaxes of the ellipse and $R_i$ the radius of the sphere having the same volume.
So the deformation shape dependence is contained in $\alpha _i $ factors. Also:

\begin{equation}
\langle i|f(\rho ,z)|j \rangle _a^b \nonumber
\end{equation}
means that the integral is performed between the limits $a$ and $b$.

\begin{equation}
\langle i|f(\rho, z) |j \rangle (z=c) \nonumber
\end{equation}
means the integrand is taken at the variable value $z=c$. This last kind of
expressions intervene when the spin-orbit 
differentiation operators act on the Heavyside $\theta (z)$
function, resulting the $\delta (z) $ function. 

We have for $\Delta V_1 $:

\begin{eqnarray}
\lefteqn{\langle n_{\rho }' m' n_z'|\Delta V_1^{(p)}|n_{\rho } m n_z \rangle = } 
\nonumber \\
 & &  \frac{1}{2}(m_0 \omega _{\rho _1}^2-m_0 \omega _1^2)
(\langle n_z'|n_z \rangle _{-\infty}^{z_{im}}+
\langle n_z' | n_z \rangle _{z_{0m}+a_m}^{\infty} \nonumber \\
 & & \cdot \frac{1}{\alpha _1^2} [(2n_{\rho }+|m|+1)\delta _{n_{\rho} ' n_{\rho }}
-\sqrt{n_{\rho }(n_{\rho }+|m|)}\delta_{n_{\rho '}n_{\rho }-1} 
 -\sqrt{(n_{\rho }+1)(n_{\rho }+|m|+1) }\delta _{n_{\rho '}n_{\rho }+1}
] \nonumber \\
 & & +\frac{1}{2}[(m_0\omega _{z_1}^2-m_0 \omega _1^2)
\langle n_z' |(z+z_1)^2|n_z \rangle _{-\infty }^{z_{im}}+
m_0\omega _{z_1}^2 \langle n_z'|(z+z_1)^2|n_z \rangle _{z_{0m}+a_m}^{\infty }
\nonumber \\
 & & -m_0\omega _2^2 \langle n_z' | (z-z_2)^2| n_z \rangle _{z_{0m}+a_m}^{\infty } ]
\delta _{n_{\rho }'n_{\rho }} \nonumber \\
 & & + \frac{1}{2}(m_0\omega _{\rho _1}^2-m_0\omega _1^2)
(\langle n_z'|n_z \rangle _{z_{im}}^0+\langle n_z'|n_z \rangle _0^{z_{0m}+a_m})
\nonumber \\
 & & \cdot \frac{1}{\alpha _1^2}
I(n_{\rho }',n_{\rho },|m|,|m|,|m|+1,|m|+1,|m|+1,\alpha _1^2 \rho _m^2(z))
\nonumber \\
 & & +\frac{1}{2}[(m_0\omega _{z_1}^2-m_0\omega _1^2)
\langle n_z'|(z+z_1)^2|n_z \rangle _{z_{im}}^0
+m_0\omega _{z_1}^2 \langle n_z'|(z+z_1)^2|n_z \rangle _0^{z_{0m}+a_m}
\nonumber \\
 & & -m_0\omega _2^2 \langle n_z' |(z-z_2)^2|n_z \rangle _0^{z_{0m}+a_m} ]
\cdot I(n_{\rho }',n_{\rho },|m|,|m|,|m|+1,|m|+1,|m|,\alpha _1^2\rho _m^2(z))
\end{eqnarray} 

For the {\it minus } terms in $\Delta V_1$:

\begin{eqnarray}
\lefteqn{\langle n_{\rho }'m'n_z' | \Delta V_1(z_i \div z_f;\rho _i \div \rho _f)|
 n_{\rho }m n_z \rangle =} \nonumber \\
 & & \frac{1}{2}m_0\omega _{\rho _1}^2 \langle n_z' |n_z \rangle _{z_i}^{z_f}
\frac{1}{\alpha _1^2}[I(n_{\rho }',n_{\rho },|m|,|m|,|m|+1,|m|+1,|m|+1,
\alpha _1^2 \rho _i ^2(z)) \nonumber \\
 & & -I(n_{\rho }',n_{\rho },|m|,|m|,|m|+1,|m|+1,|m|+1,\alpha _1^2 \rho _f^2(z))]
\nonumber \\
 & & +\frac{1}{2}m_0\omega _{z_1}^2 \langle n_z'|(z+z_1)^2|n_z \rangle _{z_i}^{z_f}
[I(n_{\rho }',n_{\rho },|m|,|m|,|m|+1,|m|+1,|m|,\alpha _1^2\rho _i^2(z)) \nonumber \\
 & & - I(n_{\rho }',n_{\rho},|m|,|m|,|m|+1,|m|+1,|m|,\alpha _1^2 \rho _f^2(z))]
\end{eqnarray}    

Matrix elements for $\Delta V_2$ reads:

\begin{eqnarray}
\lefteqn{ \langle n_{\rho }' m'n_z'|\Delta V_2^{(p)}|n_{\rho }m n_z \rangle = } 
\nonumber \\
 & & \frac{1}{2}(m_0\omega _{\rho _2}^2-m_0\omega _1^2)
(\langle n_z' |n_z \rangle _{z_{im}}^0+\langle n_z'|n_z \rangle _0^{z_{0m}+a_m}) 
\nonumber \\
 & & \cdot \frac{1}{\alpha _1^2}[(2n_{\rho }+|m|+1)\delta _{n_{\rho }'n_{\rho }}-
\sqrt{n_{\rho }(n_{\rho }+|m|)}\delta_{n_{\rho '}n_{\rho}-1}-
\sqrt{(n_{\rho }+1)(n_{\rho }+|m|+1)}\delta _{n_{\rho }'n_{\rho }+1} \nonumber \\
 & & - I(n_{\rho }',n_{\rho},|m|,|m|,|m|+1,|m|+1,|m|+1,\alpha _1^2\rho _m^2(z))] 
\nonumber \\
 & & + \frac{1}{2}[m_0\omega _{z_2}^2 \langle n_z'|(z-z_2)^2|n_z \rangle _{z_{im}}^0-
m_0\omega _1^2 \langle n_z' |(z+z_1)^2|n_z \rangle _{z_{im}}^0 \nonumber \\
 & & -(m_0\omega _{z_2}^2-m_0\omega _2^2)\langle n_z'|(z-z_2)^2|n_z \rangle
_0^{z_{0m}+a_m} ] \nonumber \\
 & & \cdot [\delta _{n_{\rho}'n_{\rho}}-I(n_{\rho }',n_{\rho },|m|,|m|,|m|+1,|m|+1,|m|,
\alpha _1^2\rho _m^2(z))]
\end{eqnarray} 
and again for the {\it minus} terms in $\Delta V_2$:

\begin{eqnarray}
\lefteqn{ \langle n_{\rho }'m'n_z'|\Delta V_2(z_i \div z_f;\rho _i \div \rho _f)|
n_{\rho }mn_z \rangle = } \nonumber \\
 & & \frac{1}{2}m_0\omega _{\rho _2}^2\langle n_z'|n_z \rangle _{z_i}^{z_f}
\frac{1}{\alpha _1^2}[I(n_{\rho }',n_{\rho },|m|,|m|,|m|+1,|m|+1,|m|+1,\alpha _1^2
\rho _i^2(z)) \nonumber \\
 & & -I(n_{\rho }',n_{\rho },|m|,|m|,|m|+1,|m|+1,|m|+1,\alpha _1^2
\rho _f^2(z)) ]\nonumber \\
 & & +\frac{1}{2}m_0\omega _{z_2}^2 \langle n_z' | (z-z_2)^2|n_z \rangle
 _{z_i}^{z_f}
 [I(n_{\rho }',n_{\rho },|m|,|m|,|m|+1,|m|+1,|m|,\alpha _1^2 \rho _i^2(z))
\nonumber \\
 & & -I(n_{\rho }',n_{\rho },|m|,|m|,|m|+1,|m|+1,|m|,\alpha _1^2 \rho _f^2(z))]
\end{eqnarray}

And finally for the neck potential operators, the expressions for the matrix elements
are:

\begin{eqnarray}
\lefteqn{ \langle n_{\rho }'m'n_z'|V_{g1}(z_i \div z_f;\rho _i \div \rho _f)|
 n_{\rho} m n_z \rangle =} \nonumber \\
 & & \left[ \left( 2V_0-\frac{m_0\omega _g^2}{2}\rho _3^2 \right )
\langle n_z'|n_z \rangle _{z_i}^{z_f}
-\frac{1}{2}m_0\omega _g^2 \langle n_z' |(z-z_3)^2|n_z \rangle _{z_i}^{z_f} \right]
\nonumber \\
 & & \cdot [I(n_{\rho }',n_{\rho },|m|,|m|,|m|+1,|m|+1,|m|,\alpha _1^2\rho _i^2(z))
\nonumber \\
 & & -\frac{1}{2}m_0\omega _g^2 \langle n_z' |n_z \rangle _{z_i}^{z_f}
\left \{ \frac{1}{\alpha _1^2}[I(n_{\rho }',n_{\rho },|m|,|m|,|m|+1,|m|+1,|m|+1,
\alpha _1^2 \rho _i^2(z)) \right.  \nonumber \\
 & & - I(n_{\rho }',n_{\rho },|m|,|m|,|m|+1,|m|+1,|m|+1,\alpha _1 ^2\rho _f^2(z))]
\nonumber \\
 & & -\frac{2 \rho _3}{\alpha _1}[I(n_{\rho }',n_{\rho },|m|,|m|,|m|+1,|m|+1,
|m|+\frac{1}{2},\alpha _1^2 \rho _i^2(z)) \nonumber \\
 & & \left. -I(n_{\rho }',n_{\rho },|m|,|m|,|m|+1,|m|+1,|m|+\frac{1}{2},\alpha _1^2
\rho _f^2(z)) ] \right \}
\end{eqnarray}
and:

\begin{eqnarray}
\lefteqn{ \langle n_{\rho }'m'n_z'|V_{g2}|n_{\rho }mn_z \rangle =} \nonumber \\
 & & V_0 \{ \langle n_z'|n_z \rangle _{z_{c1}}^{z_s}
[I(n_{\rho }',n_{\rho },|m|,|m|,|m|+1,|m|+1,|m|,\alpha _1 ^2\rho _1^2(z)) 
\nonumber \\
 & & -I(n_{\rho }',n_{\rho },|m|,|m|,|m|+1,|m|+1,|m|,\alpha _1^2\rho _g^2(z))]
\nonumber \\
 & & +\langle n_z'|n_z \rangle _{z_s}^{z_{c2}}
[I(n_{\rho }',n_{\rho },|m|,|m|,|m|+1,|m|+1,|m|,\alpha _1^2 \rho _2^2(z)) 
\nonumber \\
 & & -I(n_{\rho }',n_{\rho },|m|,|m|,|m|+1,|m|+1,|m|,\alpha _1^2\rho _g^2(z))] \}
\end{eqnarray}

For the $z$-dependent terms of the matrix elements, the following abreviations
are available \cite{bad} :

\begin{eqnarray}
j_{\nu _1,\nu _2,\zeta _i} & = &
\int _0^{\infty } d\zeta e^{-(\zeta -\zeta _i)^2} \mathcal{H}_{\nu _1}
(\zeta -\zeta _i)\mathcal{H}_{\nu _2}(\zeta -\zeta _i) \nonumber \\
& & \int _{-\infty }^0 e^{-(\zeta +\zeta _i)^2} \mathcal{H}_{\nu _1}
[-(\zeta +\zeta _i)]\mathcal{H}_{\nu _2}[-(\zeta +\zeta _i)] d\zeta
\end{eqnarray}

\begin{equation}
j_{\nu _1,\nu _2,\zeta _i-\zeta _0}=\int _{\zeta _0}^{\infty }
e^{-(\zeta - \zeta _i)^2}\mathcal{H}_{\nu _1}(\zeta -\zeta _i)
\mathcal{H}_{\nu _2}(\zeta -\zeta _i)d\zeta
\end{equation}
where the general formula for computing these kind of intergals is:

\begin{equation}
j_{\nu _1,\nu _2,\zeta _0}=
\frac{e^{-\zeta _0^2}}{\nu _1 -\nu _2}
[\nu _1 \mathcal{H}_{\nu _1-1}(-\zeta _0)\mathcal{H}_{\nu _2}(-\zeta _0)
-\nu _2 \mathcal{H}_{\nu _1}(-\zeta _0)\mathcal{H}_{\nu _2 -1}(-\zeta _0) ]
\end{equation}
and $\mathcal{H}_{\nu}(\zeta)$ is the Hermite function.

\begin{eqnarray}
\langle n_z' | n_z \rangle _{z_i}^{z_f}= 
  & C_{\nu _1 '}C_{\nu _1}\frac{1}{\alpha _1}
[j_{\nu _1',\nu _1,\alpha (z_1+z_f)}-j_{\nu _1',\nu _1, \alpha _1(z_1+z_i)}]
 & ,z_i,z_f<0 \nonumber \\
 = & C_{\nu _2'}C_{\nu _2}\frac{1}{\alpha _2}
[j_{\nu _2',\nu _2,\alpha _2(z_2-z_i)}-
j_{\nu _2',\nu _2,\alpha _2(z_2-z_f)}]
 & ,z_i,z_f \ge 0
\end{eqnarray}

\begin{eqnarray}
\langle n_z'|(z+z_1)^2|n_z \rangle _{z_i}^{z_f} & = 
C_{\nu _1'}C_{\nu _1 }\frac{1}{\alpha _1^3} &
\left \{ \frac{1}{4}[j_{\nu _1'+1,\nu _1+1,\alpha _1(z_1+z_f)}
-j_{\nu _1'+1,\nu _1+1,\alpha _1(z_1+z_i)}] \right. \nonumber \\
 & &+ \frac{\nu _1'}{2}[j_{\nu _1'+1,\nu _1-1,\alpha _1(z_1+z_f)}
-j_{\nu _1'+1,\nu _1-1,\alpha _1(z_1+z_i)}] \nonumber \\
 & & + \frac{\nu _1}{2}[j_{\nu _1'-1,\nu _1+1,\alpha _1(z_1+z_f)}
-j_{\nu _1'-1,\nu _1+1,\alpha _1(z_1+z_i)}] \nonumber \\
 & & \left. +
\nu _1'\nu _1[j_{\nu _1'-1,\nu _1-1,\alpha _1(z_1+z_f)}-
j_{\nu _1'-1,\nu _1-1,\alpha _1(z_1+z_i)}] \right \}
\end{eqnarray}

\begin{eqnarray}
\langle n_z' |(z-z_2)^2|n_z \rangle _{z_i}^{z_f} & =
C_{\nu _2'}C_{\nu _2}\frac{1}{\alpha _2^3} &
\left \{ \frac{1}{4}[j_{\nu _2'+1,\nu _2+1,\alpha _2(z_2-z_i)}
-j_{\nu _2'+1,\nu _2+1,\alpha _2(z_2-z_f)}]  \right. \nonumber \\
 & & + \frac{\nu _2}{2}[j_{\nu _2'+1,\nu _2-1,\alpha _2(z_2-z_i)}
-j_{\nu _2'+1,\nu _2-1,\alpha _2(z_2-z_f)}] \nonumber \\
 & & +\frac{\nu _2'}{2}[j_{\nu _2'-1,\nu _2+1,\alpha _2(z_2-z_i)}
-j_{\nu _2'-1,\nu _2+1,\alpha _2(z-z_f)}] \nonumber \\
 & & \left. +\nu _2'\nu _2[j_{\nu _2'-1,\nu _2-1,\alpha _2(z_2-z_i)}
-j_{\nu _2'-1,\nu _2-1,\alpha _2(z_2-z_f)}] \right \}
\end{eqnarray}

\begin{eqnarray}
\langle n_z' |z+z_1 | n_z \rangle _{z_i}^{z_f} & = C_{\nu _1'}
C_{\nu _1}\frac{1}{\alpha _1^2} &
\displaystyle \left \{ \frac{1}{2}[j_{\nu _1',\nu _1+1,\alpha _1(z_1+z_i)}
 -j_{\nu _1',\nu _1+1,\alpha _1(z_1+z_f)}] \right. \nonumber \\
 & & \displaystyle \left. + \nu _1[j_{\nu _1',\nu _1-1,\alpha _1(z_1+z_i)}
-j_{\nu _1',\nu _1-1,\alpha _1(z_1+z_f)}] \right \}
\end{eqnarray}

\begin{eqnarray}
\langle n_z'| z-z_2 | n_z \rangle _{z_i}^{z_f} & = C_{\nu _2'}
C_{\nu _2}\frac{1}{\alpha _2^2} &
\left \{ \frac{1}{2}[j_{\nu _2',\nu _2+1,\alpha _2(z_2-z_i)}
 -j_{\nu _2',\nu _2+1,\alpha _2(z_2-z_f)}] \right. \nonumber \\
 & & \left. + \nu _2[j_{\nu _2',\nu _2-1,\alpha _2(z_2-z_i)}
 -j_{\nu _2',\nu _2-1,\alpha _2(z_2-z_f)}] \right. \}
\end{eqnarray} 

For $(z-z_3)$ dependent terms, one can use:

\begin{equation}
\langle n_z'|(z-z_3)^2| n_z \rangle _{z_i}^{z_f}=
 \langle n_z'|(z-z_2)^2|n_z \rangle _{z_i}^{z_f}
+2(z_2-z_3) \langle n_z'|z-z_2|n_z \rangle _{z_i}^{z_f}
+(z_2-z_3)^2 \langle n_z'| n_z \rangle _{z_i}^{z_f}
\end{equation}

\begin{eqnarray}
\lefteqn{ \Big \langle n_z' \Big|\frac{\partial }{\partial z} \Big |
 n_z \Big \rangle _{z_i}^{z_f} =} \nonumber \\
 =   C_{\nu _1'}C_{\nu _1} & \left [ \frac{1}{2}(j_{\nu _1',\nu _1+1,\alpha _1
(z_1+z_f)}-j_{\nu _1',\nu _1+1,\alpha _1(z_1+z_i)})  \right. &  \nonumber \\
   &  \left. +\nu _1(j_{\nu _1',\nu _1-1,\alpha _1(z_1+z_f)}
-j_{\nu _1',\nu _1-1,\alpha _1(z_1+z_i)}) \right ] & z_i,z_f<0 \nonumber \\
 = C_{\nu _2'}C_{\nu _2} & \left [ - \frac{1}{2}(j_{\nu _2',\nu _2+1,\alpha _2
(z_2-z_i)}-j_{\nu _2',\nu '2+1,\alpha _2(z_2-z_f)}) \right. & \nonumber \\
 & \left. +\nu _2(j_{\nu _2',\nu _2-1,\alpha _2(z_2-z_i)}
-j_{\nu _2',\nu _2-1,\alpha _2(z_2-z_f)}) \right ] & z_i,z_f \ge 0
\end{eqnarray}

\section{Spin-orbit and $\mathbf \Omega ^2$ matrix elements}

The necessary expressions to calculate the matrix elements for spin-orbit
and $\mathbf \Omega ^2$ terms are given in this appendix.

For the $(v_1)$ volume, we use the abreviation:

\begin{eqnarray}
w_1(v_1) & = & (\sqrt{n_{\rho }+1}\delta _{n_{\rho '}n_{\rho }+1}+
\sqrt{n_{\rho }+|m|}\delta _{n_{\rho '}n_{\rho }})
2m_0\omega _{z_1}^2 \alpha _1 \nonumber \\
 & & \cdot (\langle n_z'|z+z_1|n_z \rangle _{-\infty}^{z_{im}}+
\langle n_z' |z+z_1|n_z \rangle _{z_{0m}+a_m}^{\infty } \nonumber \\
 & & +(\sqrt{n_{\rho }+1}\delta _{n_{\rho '}n_{\rho }+1}-
\sqrt{n_{\rho }+|m|}\delta _{n_{\rho }'n_{\rho }})\cdot 
\frac{m_0\omega _{\rho _1}^2}{\alpha _1} \nonumber \\
 & & \cdot \left [2 \left ( \Big \langle n_z' \Big |\frac{\partial }{\partial z} 
\Big | n_z \Big \rangle _{-\infty }^{z_{im}}+
\Big \langle n_z' \Big | \frac{\partial }{\partial z} \Big | n_z \Big \rangle
_{z_{0m}+a_m}^{\infty} \right ) \right. \nonumber \\
 & & + \langle n_z'|n_z \rangle (z=z_{0m}+a_m)
-\langle n_z'| n_z \rangle (z=z_{im}) \Big ]
\end{eqnarray}

\begin{eqnarray}
w_2(v_1) & = & (\sqrt{n_{\rho }+|m|+1}\delta _{n_{\rho }'n_{\rho }}
+\sqrt{n_{\rho }}\delta _{n_{\rho }'n_{\rho }-1})\cdot
2m_0\omega_{z_1}^2\alpha _1 \nonumber \\
 & & \cdot (\langle n_z'|z+z_1|n_z \rangle _{-\infty }^{z_{im}}
+\langle n_z' | z+z_1|n_z \rangle _{z_{0m}+a_m}^{\infty}) \nonumber \\
 & & +(\sqrt{n_{\rho }+|m|+1}\delta _{n_{\rho }'n_{\rho }}
-\sqrt{n_{\rho }}\delta _{n_{\rho }'n_{\rho }-1}) \cdot
\frac{m_0\omega _{\rho _1}^2}{\alpha _1} \nonumber \\
 & & \cdot \Big [ 2 \Big ( \Big \langle n_z' \Big | \frac{\partial }{\partial z}
\Big | n_z \Big \rangle _{-\infty }^{z_{im}}+
\Big \langle n_z'\Big | \frac{\partial }{\partial z} \Big | n_z
\Big \rangle _{z_{0m}+a_m}^{\infty } \nonumber \\
 & & +\langle n_z'|\rangle n_z (z=z_{0m}+a_m)-
\langle n_z'|n_z \rangle (z=z_{im}) ]
\end{eqnarray} 

Then we have:

\begin{equation}
\begin{array}{ll}
\langle n_{\rho }'m'n_z'|\{ \mathbf \Omega^+(v_1),[1-\theta (z-z_{im}) \}
+ \{ \mathbf \Omega ^+(v_1),[z-(z_{0m}+a_m)] \}
|n_{\rho }mn_z \rangle &  
  \nonumber \\
  \qquad \qquad  \qquad = \delta _{m' m+1}\cdot w_1(v_1)   \qquad ,m<0 &   \nonumber \\
   \qquad  \qquad  \qquad = \delta _{m'm+1}\cdot (-w_2(v_1))  \qquad ,m \ge 0 & 
\end{array}
\end{equation}

\begin{equation}
\begin{array}{ll}
\langle n_{\rho }'m'n_z'|\{ \mathbf \Omega^-(v_1),[1-\theta (z-z_{im}) \}
+ \{ \mathbf \Omega ^+(v_1),[z-(z_{0m}+a_m)] \} |n_{\rho }m n_z \rangle &
\nonumber \\
\qquad \qquad \qquad = \delta _{m'm-1}\cdot w_2(v_1) \qquad ,m \le 0 & \nonumber \\
\qquad \qquad \qquad = \delta _{m'm-1}\cdot (-w_1(v_1)) \qquad ,m>0 &
\end{array}
\end{equation}

Because of the symmetry we have:

\begin{equation}
\langle '|\{\mathbf \Omega ^+,f(\rho ,z) \}| \rangle (m',m)=
\langle | \{ \mathbf \Omega ^-,f(\rho ,z) \} | ' \rangle (m,m')
\end{equation}
So, only one half of the $\mathbf \Omega \mathbf s$ and $\mathbf \Omega ^2$
matrix elements must be calculated. Further on, only the $\mathbf \Omega ^+$
matrix elements are given, the $\mathbf \Omega ^-$ part resulting from symmetry.

\begin{eqnarray}
\lefteqn{
\langle n_{\rho }'m'n_z' |\{ \mathbf \Omega ^+(v_1),
[\theta (z-z_{im})-\theta [z-(z_{0m}+a_m)]]\theta (\rho -\rho _m(z))\}|n_{\rho }
mn_z \rangle=} \nonumber \\
& & \delta _{m'm+1}\{-I(n_{\rho }',n_{\rho },|m|-1,|m|,|m|,|m|+1,|m|,
\alpha _1^2\rho _m^2(z)) \nonumber \\
& & \cdot \Big [ 2m_0\omega _{z_1}^2\alpha _1 \langle n_z'|z+z_1|n_z \rangle
_{z_{im}}^{z_{0m}+a_m} \nonumber \\
 & & +\frac{m_0\omega _{\rho _1}^2}{\alpha _1}
\Big (2\Big \langle n_z'\Big | \frac{\partial }{\partial z} \Big |n_z
\Big \rangle_{z_{im}}^{z_{0m}+a_m}
+\langle n_z'|n_z \rangle (z=z_{im})
-\langle n_z'|n_z \rangle (z=z_{0m}+a_m) \Big ) \Big ] \nonumber \\
 & & +2m_0\omega _{z_1}^2\alpha _1 \langle n_z'|z+z_1|n_z \rangle _{z_{im}}^{
z_{0m}+a_m} \nonumber \\
 & & \cdot [2(n_{\rho }+|m|)
I(n_{\rho }',n_{\rho },|m|-1,|m|-1,|m|,|m|+1,|m|-1,\alpha _1^2\rho _m^2(z))
\nonumber \\
 & & +\frac{1}{2\alpha _1}G(n_{\rho }',n_{\rho },|m|-1,|m|,\alpha _1^2
\rho _m^2(z))] \} \qquad ,m<0 \\
\end{eqnarray} 
and:

\begin{eqnarray}
\lefteqn{
\langle n_{\rho }'m'n_z' |\{ \mathbf \Omega ^+(v_1),
[\theta (z-z_{im})-\theta [z-(z_{0m}+a_m)]]\theta (\rho -\rho _m(z))\}|n_{\rho }
mn_z \rangle=} \nonumber \\
 & & \delta _{m'm+1}\{I(n_{\rho }',n_{\rho },|m|+1,|m|,|m|+2,|m|+1,
|m|+1,\alpha _1^2 \rho _m^2(z)) \nonumber \\
 & & \cdot \Big [2m_0 \omega _{z_1}^2\alpha _1 \langle n_z'|z+z_1|n_z \rangle
_{z_{im}}^{z_{0m}+a_m} \nonumber \\
 & &-\frac{m_0 \omega _{\rho _1}^2}{\alpha _1}
\Big (2\Big \langle n_z'\Big |\frac{\partial }{\partial z} \Big |
n_z \Big \rangle _{z_{im}}^{z_{0m}+a_m}
+\langle n_z'|n_z \rangle (z=z_{im})-\langle n_z'|n_z \rangle (z=z_{0m}+a_m)
\Big ) \Big ] \nonumber \\
 & & -2m_0 \omega _{z_1}^2\alpha _1 \langle n_z' | z+z_1 |n_z
\rangle _{z_{im}}^{z_{0m}+a_m} \nonumber \\
 & & \cdot [2I(n_{\rho }',n_{\rho },|m|+1,|m|+1,|m|+2,|m|+1,|m|+1,
\alpha _1^2\rho _m^2(z)) \nonumber \\
 & &\frac{1}{2\alpha _1}G(n_{\rho }',n_{\rho },|m|+1,|m|,\alpha _1^2
\rho _m^2(z)) ] \} \qquad ,m \ge 0
\end{eqnarray}  

\begin{eqnarray}
\lefteqn{
\langle n_{\rho }'m'n_z' |\{ \mathbf \Omega ^+(v_1),
[\theta (z-z_i)-\theta (z-z_f)][\theta (\rho -\rho _i(z))
-\theta (\rho -\rho _f(z))] \} | n_{\rho }mn_z \rangle =} \nonumber \\
& & \delta _{m'm+1} \{ [-I(n_{\rho }',n_{\rho },|m|-1,|m|,|m|,|m|+1,|m|,
\alpha _1^2\rho _i^2(z) \nonumber \\
 & & +I(n_{\rho }',n_{\rho },|m|-1,|m|,|m|,|m|+1,|m|,\alpha _1^2\rho _f^2(z))]
\nonumber \\
 & & \cdot \Big [ 2m_0\omega _{z_1}^2\alpha _1 \langle n_z'|z+z_1|
\rangle _{z_i}^{z_f} \nonumber \\
 & & +\frac{m_0\omega _{\rho _1}^2}{\alpha _1}
\Big ( 2\Big \langle n_z' \Big | \frac{\partial }{\partial z} \Big|
n_z \Big \rangle _{z_i}^{z_f} +\langle n_z' |n_z \rangle (z=z_i)
-\langle n_z'| n_z \rangle (z=z_f) \Big ) \Big ] \nonumber \\
 & & +2m_0\omega _{z_1}^2\alpha _1 \langle n_z'|z+z_1|n_z \rangle _{z_i}^{z_f}
\nonumber \\
 & & \cdot [2(n_{\rho }+|m|)[I(n_{\rho }',n_{\rho },|m|-1,|m|-1,|m|,|m|+1,|m|-1,
\alpha _1^2\rho _i^2(z)) \nonumber \\
 & & -I(n_{\rho }',n_{\rho },|m|-1,|m|-1,|m|,|m|+1,|m|-1,\alpha _1^2\rho _f^2(z))]
\nonumber \\
 & & +\frac{1}{2\alpha _1}[G(n_{\rho }',n_{\rho },|m|-1,|m|,\alpha _1^2 \rho _1^2(z))
-G(n_{\rho }',n_{\rho },|m|-1,|m|,\alpha _1^2\rho _f^2(z))]] \} 
\nonumber \\
 & & \qquad \qquad \qquad \qquad \qquad \qquad \qquad \qquad \qquad \qquad \qquad  ,m<0
\end{eqnarray}  
and the same term for $m \ge 0 $ :

\begin{eqnarray}
\lefteqn{
\langle n_{\rho }'m'n_z' |\{ \mathbf \Omega ^+(v_1),
[\theta (z-z_i)-\theta (z-z_f)][\theta (\rho -\rho _i(z))
-\theta (\rho -\rho _f(z))] \} | n_{\rho }mn_z \rangle =} \nonumber \\
 & & \delta _{m'm+1} \{ [I(n_{\rho }',n_{\rho },|m|+1,|m|,|m|+2,|m|+1,|m|+1,
\alpha _1^2 \rho _i^2(z)) \nonumber \\
 & & -I(n_{\rho }',n_{\rho },|m|+1,|m|,|m|+2,|m|+1,|m|+1,\alpha _1^2 \rho _f^2(z))]
\nonumber \\
 & & \cdot \Big [2m_0\omega _{z_1}^2\alpha _1 \langle n_z'|z+z_1|n_z
\rangle _{z_i}^{z_f} \nonumber \\
 & & -\frac{m_0\omega _{\rho _1}^2}{\alpha _1}
\Big (2 \Big \langle n_z' \Big | \frac{\partial }{\partial z} \Big | n_z 
\Big \rangle _{z_i}^{z_f}+ \langle n_z' |n_z \rangle (z=z_i)
-\langle n_z'|n_z \rangle (z=z_f) \Big ) \Big ] \nonumber \\
 & & -2m_0\omega _{z_1}^2\alpha _1 \langle n_z'|z+z_1 |n_z \rangle _{z_i}^{z_f}
\nonumber \\
 & & \cdot [2(I(n_{\rho }',n_{\rho },|m|+1,|m|+1,|m|+2,|m|+1,|m|+1,\alpha _1^2
\rho _i^2(z)) \nonumber \\
 & & -I(n_{\rho }',n_{\rho },|m|+1,|m|+1,|m|+2,|m|+1,|m|+1,\alpha _1^2\rho _f^2(z))
\nonumber \\
 & & -\frac{1}{2\alpha _1}(G(n_{\rho }',n_{\rho },|m|+1,|m|,\alpha _1^2\rho _i^2(z))
-G(n_{\rho }',n_{\rho },|m|+1,|m|,\alpha _1^2\rho _f^2(z)) ] \}
\nonumber \\
& & \qquad \qquad \qquad \qquad \qquad \qquad \qquad \qquad \qquad \qquad \qquad ,
m \ge 0
\end{eqnarray}

\begin{equation}
\langle n_{\rho }' m'n_z'|\mathbf \Omega _z(v_1) |n_{\rho }mn_z \rangle
=m_0\omega _{\rho _1}^2 m \delta _{m'm}
\end{equation}

We now present the $(v_2)$ -dependent necessary terms for $\mathbf \Omega
\mathbf s (v_2)$ matrix elements construction: 

\begin{eqnarray}
\lefteqn{
\langle
n_{\rho }'m'n_z' | \{ \mathbf \Omega ^+(v_2),[\theta (z-z_{im})-
\theta [z-(z_{0m}+a_m)]][1-\theta (\rho -\rho _m(z)) ] \} |n_{\rho }mn_z
\rangle =} \nonumber \\
 & & \delta _{m'm+1} \Big \{ 2m_0\omega _{z_2}^2 \alpha _1
\langle n_z'|z-z_2|n_z \rangle _{z_{im}}^{z_{0m}+a_m}
(\sqrt{n_{\rho }+|m|}\delta _{n_{\rho }'n_{\rho }}+\sqrt{n_{\rho}+1}
\delta _{n_{\rho }'n_{\rho }+1} ) \nonumber \\
 & & -\frac{m_0\omega _{\rho _2}^2}{2}
\Big (2 \Big \langle n_z' \Big | \frac{\partial }{\partial z} |n_z
\Big \rangle _ {z_{im}}^{z_{0m}+a_m} +\langle n_z'|n_z \rangle (z=z_{im})
-\langle n_z'|n_z \rangle (z=z_{0m}+a_m) \Big ) \nonumber \\
 & & \cdot ( \sqrt{n_{\rho }+|m|}\delta _{n_{\rho }'n_{\rho }}
-\sqrt{n_{\rho }+1}\delta _{n_{\rho }'n_{\rho }+1}) \nonumber \\
 & & +I(n_{\rho}',n_{\rho},|m|-1,|m|,|m|,|m|+1,|m|,\alpha_1^2\rho_m^2(z))
\nonumber \\
 & & \cdot \Big [ 2m_0\omega _{z_2}^2\alpha _1 \langle n_z'|z-z_2|n_z \rangle
_{z_{im}}^{z_{0m}+a_m} \nonumber \\
 & & + \frac{m_0\omega _{\rho _2}^2}{\alpha _1 }
\Big (2 \Big \langle n_z' \Big | \frac {\partial }{\partial z} \Big | n_z
\Big \rangle _{z_{im}}^{z_{0m}+a_m}
+\langle n_z'|n_z \rangle (z=z_{im})-\langle n_z'|n_z \rangle (z=z_{0m}+a_m)
\Big ) \Big ] \nonumber \\
 & & -2m_0 \omega _{z_2}^2 \alpha _1 \langle n_z' | z-z_2|n_z \rangle _{z_{im}}
^{z_{0m}+a_m} \nonumber \\
 & & \cdot [ 2(n_{\rho }+|m|)I(n_{\rho }',n_{\rho },|m|-1,|m|-1,|m|,|m|+1,|m|-1,
\alpha_1^2\rho _m^2(z)) \nonumber \\
 & & +\frac{1}{2\alpha _1}G(n_{\rho }',n_{\rho },|m|-1,|m|,\alpha _1^2\rho _m^2(z))
] \Big \} \qquad ,m<0
\end{eqnarray}

\begin{eqnarray}
\lefteqn{
\langle
n_{\rho }'m'n_z' | \{ \mathbf \Omega ^+(v_2),[\theta (z-z_{im})-
\theta [z-(z_{0m}+a_m)]][1-\theta (\rho -\rho _m(z)) ] \} |n_{\rho }mn_z
\rangle =} \nonumber \\
 & & \delta _{m'm+1} \Big \{ -2m_0\omega _{z_2}^2\alpha _1
\langle n_z'|z-z_2 | n_z \rangle _{z_{im}}^{z_{0m}+a_m}
(\sqrt{n_{\rho }+|m|+1} \delta _{n_{\rho }'n_{\rho}}+\sqrt{n_{\rho }}
\delta _{n_{\rho }'n_{\rho }-1}) \nonumber \\
 & & - \frac{m_0\omega _{\rho _2}^2}{\alpha _1}
\Big (2 \Big \langle n_z' \Big | \frac{\partial }{\partial z} \Big |n_z
\Big \rangle _{z_{im}}^{z_{0m}+a_m}+\langle n_z'|n_z \rangle (z=z_{im})
-\langle n_z'|n_z \rangle (z=z_{0m}+a_m) \Big ) \nonumber \\
 & & \cdot ( \sqrt{n_{\rho }+|m|+1}\delta _{n_{\rho }'n_{\rho }}
-\sqrt{n_{\rho }}\delta _{n_{\rho }'n_{\rho }-1} ) \nonumber \\
 & & -I(n_{\rho }',n_{\rho },|m|+1,|m|,|m|+2,|m|+1,|m|+1,\alpha _1^2\rho _m^2(z))
\nonumber \\
 & & \cdot \Big [ 2m_0\omega _{z_2}^2\alpha _1 \langle n_z'|z-z_2|n_z \rangle
_{z_{im}}^{z_{0m}+a_m} \nonumber \\
 & & - \frac{m_0\omega _{\rho _2}^2}{\alpha _1} \Big ( 
2\Big \langle n_z' \Big | \frac{\partial }{\partial z} \Big | n_z \Big \rangle
_{z_{im}}^{z_{0m}+a_m}+\langle n_z'|n_z \rangle (z=z_{im})-
\langle n_z'|n_z \rangle (z=z_{0m}+a_m) \Big ) \Big ] \nonumber \\
 & &+ 2m_0\omega _{z_2}^2\alpha _1 \langle n_z'|z-z_2| n_z \rangle _{z_{im}}
^{z_{0m}+a_m} \nonumber \\
 & & \cdot [2I(n_{\rho }',n_{\rho },|m|+1,|m|+1,|m|+2,|m|+1,|m|+1,\alpha _1^2
\rho _m^2(z)) \nonumber \\
 & & -\frac{1}{2\alpha _1}G(n_{\rho }',n_{\rho },|m|+1,|m|,\alpha _1 \rho_m^2(z))
\Big ] \Big \} \qquad ,m \ge 0
\end{eqnarray}

\begin{eqnarray}
\lefteqn{
\langle n_{\rho }'m'n_z'| \{ \mathbf \Omega ^+(v_2),
[\theta (z-z_i)-\theta (z-z_f)]
[\theta (\rho -\rho _i(z))\theta (\rho -\rho _f(z))] \} |n_{\rho }mn_z 
\rangle =} \nonumber \\
& & \delta _{m'm+1} \Big \{ [-I(n_{\rho }',n_{\rho },|m|-1,|m|,|m|,|m|+1,
|m|,\alpha _1^2\rho _i^2(z)) \nonumber \\
 & & +I(n_{\rho}',n_{\rho },|m|-1,|m|,|m|,|m|+1,|m|,\alpha _1^2\rho _f^2(z))]
\nonumber \\
& & \cdot \Big [2m_0\omega _{z_2}^2\alpha _1 \langle n_z'|z-z_2|n_z \rangle
_{z_i}^{z_f} \nonumber \\
 & & + \frac{m_0\omega _{\rho _2}^2}{\alpha _1}
\Big ( 2 \Big \langle n_z' \Big | \frac{\partial }{\partial z} \Big |
n_z \Big \rangle _{z_i}^{z_f}+\langle n_z' | n_z \rangle (z=z_i)-
\langle n_z'|n_z \rangle (z=z_f) \Big ) \Big ] \nonumber \\
 & & +2m_0\omega_{z_2}^2\alpha _1 \langle n_z'|z-z_2|n_z \rangle _{z_i}^{z_f}
\nonumber \\
 & & \cdot [2(n_{\rho }+|m|)(I(n_{\rho }',n_{\rho },|m|-1,|m|-1,|m|,|m|+1,
|m|-1,\alpha _1^2\rho _i^2(z)) \nonumber \\
 & & -I(n_{\rho }',n_{\rho },|m|-1,|m|-1,|m|,|m|+1,|m|-1,\alpha _1^2\rho _f^2(z)))
\nonumber \\
 & & +\frac{1}{2\alpha _1}(G(n_{\rho }',n_{\rho },|m|-1,|m|,
\alpha _1^2\rho _i^2(z))-
G(n_{\rho }',n_{\rho },|m|-1,|m|,\alpha_1^2\rho _f^2(z)) ) ] \Big \}
\nonumber \\
& & \qquad \qquad \qquad \qquad \qquad \qquad \qquad \qquad ,m<0
\end{eqnarray}

\begin{eqnarray}
\lefteqn{
\langle n_{\rho }'m'n_z'| \{ \mathbf \Omega ^+(v_2),
[\theta (z-z_i)-\theta (z-z_f)]
[\theta (\rho -\rho _i(z))\theta (\rho -\rho _f(z))] \} |n_{\rho }mn_z 
\rangle =} \nonumber \\
 & & \delta _{m'm+1} \Big \{ [I(n_{\rho }',n_{\rho },|m|+1,|m|,|m|+2,
|m|+1,|m|+1,\alpha_1^2\rho _i^2(z)) \nonumber \\
 & & -I(n_{\rho }',n_{\rho },|m|+1,|m|,|m|+2,|m|+2,|m|+1,|m|+1,
\alpha _1^2\rho _f^2(z)) ] \nonumber \\
 & & \cdot \Big [ 2m_0\omega _{z_2}^2\alpha _1\langle n_z'|z-z_2|n_z \rangle
_{z_i}^{z_f} \nonumber \\
 & & -\frac{m_0\omega _{\rho _2}^2}{\alpha _1}
\Big (2 \Big \langle n_z' \Big | \frac{\partial }{\partial z} \Big | n_z
\Big \rangle _{z_i}^{z_f} +\langle n_z'|n_z \rangle (z=z_i)
-\langle n_z' |n_z \rangle (z=z_f) \Big ) \Big ] \nonumber \\
 & & -2m_0\omega _{z_2}^2\alpha _1 \langle n_z'|z-z_2|n_z \rangle
_{z_i}^{z_f} \nonumber \\
& & \cdot [2(I(n_{\rho }',n_{\rho },|m|+1,|m|+1,|m|+2,|m|+1,|m|+1,
\alpha _1^2\rho _i^2(z)) \nonumber \\
 & &-I(n_{\rho }',n_{\rho },|m|+1,|m|+1,|m|+2,|m|+1,|m|+1,\alpha_1^2\rho _f^2
(z))) \nonumber \\
 & & -\frac{1}{2\alpha _1}(G(n_{\rho }',n_{\rho },|m|+1,|m|,\alpha _1^2\rho _i^2
(z))-G(n_{\rho}',n_{\rho },|m|+1,|m|,\alpha _1^2\rho _f^2(z))) ] \Big \}
\nonumber \\
& & \qquad \qquad \qquad \qquad \qquad \qquad \qquad \qquad ,m \ge 0
\end{eqnarray}

The diagonal term generated by $\mathbf \Omega _z $ is:

\begin{equation}
\langle n_{\rho }'m'n_z'|\mathbf \Omega _z (v_2)|n_{\rho }mn_z
\rangle =m_0\omega _{\rho _2}^2m\delta _{m'm}
\end{equation}

For the neck region elements we have:

\begin{eqnarray}
\lefteqn{
\langle n_{\rho }'m'n_z' |\{ \mathbf \Omega ^+(v_{g1}),
[\theta (z-z_i)-\theta (z-z_f)][\theta (\rho -\rho _i(z))-
\theta (\rho -\rho _f(z))] \} |n_{\rho }mn_z \rangle =} 
\nonumber \\
& & \delta_{m'm+1}m_0\omega _g^2 \Big \{ \{\frac{1}{\alpha _1} 
[I(n_{\rho }',n_{\rho },|m|-1,|m|,|m|,|m|+1,|m|,\alpha_1^2\rho_i^2(z))
\nonumber \\
 & & -I(n_{\rho }',n_{\rho },|m|-1,|m|,|m|,|m|+1,|m|,\alpha _1^2\rho_f^2(z))]
\nonumber \\
 & & -\rho _3[I(n_{\rho }',n_{\rho },|m|-1,|m|,|m|,|m|+1,|m|-\frac{1}{2},
\alpha_1^2\rho_i^2(z)) \nonumber \\
 & & -I(n_{\rho }',n_{\rho },|m|-1,|m|,|m|,|m|+1,|m|-\frac{1}{2},
\alpha _1^2\rho_f^2(z))] \} \nonumber \\
 & & \cdot \Big [2\Big \langle n_z' \Big | \frac{\partial }{\partial z} 
\Big |n_z \Big \rangle _{z_i}^{z_f}+\langle n_z'|n_z \rangle (z=z_i)
-\langle n_z'|n_z \rangle (z=z_f) \Big ] \nonumber \\
 & & -2 \{\alpha _1 \{2(n_{\rho}+|m|)[
I(n_{\rho }',n_{\rho },|m|-1,|m|-1,|m|,|m|+1,|m|-1,\alpha_1^2\rho_i^2(z))
\nonumber \\
 & & -I(n_{\rho }',n_{\rho },|m|-1,|m|-1,|m|,|m|+1,|m|-1,\alpha_1^2\rho_f^2(z))]
\nonumber \\
 & &-[I(n_{\rho }',n_{\rho },|m|-1,|m|,|m|,|m|+1,|m|,\alpha _1^2\rho _i^2(z))
\nonumber \\
 & & -I(n_{\rho}',n_{\rho },|m|-1,|m|,|m|,|m|+1,|m|,\alpha _1^2\rho_f^2(z))] \}
\nonumber \\
& & +\frac{1}{2}[G(n_{\rho }',n_{\rho },|m|-1,|m|,\alpha_1^2\rho _i^2(z))-
G(n_{\rho }',n_{\rho },|m|-1,|m|,\alpha_1^2\rho_f^2(z))]
\nonumber \\
 & & \cdot \langle n_z'|z-z_3|n_z \rangle _{z_i}^{z_f} \} \Big \}
\qquad ,m<0
\end{eqnarray}

\begin{eqnarray}
\lefteqn{
\langle n_{\rho }'m'n_z' |\{ \mathbf \Omega ^+(v_{g1}),
[\theta (z-z_i)-\theta (z-z_f)][\theta (\rho -\rho _i(z))-
\theta (\rho -\rho _f(z))] \} |n_{\rho }mn_z \rangle =} 
\nonumber \\
& & \delta_{m'm+1}m_0\omega _g^2\Big \{ [
I(n_{\rho }',n_{\rho },|m|+1,|m|,|m|+2,|m|+1,|m|+1,\alpha_1^2\rho_i^2(z))
\nonumber \\
 & & -I(n_{\rho }',n_{\rho },|m|+1,|m|,|m|+2,|m|+1,|m|+1,\alpha_1^2\rho_f^2
(z)) ] \nonumber \\
 & & \cdot \Big \{ \frac{1}{\alpha_1} \Big [2 \Big \langle n_z' \Big |
\frac{\partial }{\partial z} \Big |n_z \Big \rangle _{z_i}^{z_f}
+\langle n_z'|n_z \rangle (z=z_i)-\langle n_z'|n_z \rangle (z=z_f)
\Big ] \nonumber \\
 & & -2\alpha _ \langle n_z'|z-z_3|n_z \rangle _{z_i}^{z_f} \Big \}
\nonumber \\
 & &-\rho_3 [I(n_{\rho }',n_{\rho },|m|+1,|m|,|m|+2,|m|+1,|m|+\frac{1}{2},
\alpha _1^2\rho _i^2(z)) \nonumber \\
 & & -I(n_{\rho }',n_{\rho },|m|+1,|m|,|m|+2,|m|+1,|m|+\frac{1}{2},
\alpha_1^2\rho_f^2(z))] \nonumber \\
 & & \cdot \Big [2 \Big \langle n_z'\Big | \frac{\partial }{\partial z}
\Big | n_z \Big \rangle _{z_i}^{z_f}+\langle n_z'|n_z \rangle(z=z_i)
-\langle n_z'|n_z \rangle (z=z_f) \Big ] \nonumber \\
& & +2\{ 2\alpha_1[I(n_{\rho}',n_{\rho },|m|+1,|m|+1,|m|+2,|m|+1,|m|+1,
\alpha_1^2\rho_i^2(z)) \nonumber \\
 & & -I(n_{\rho }',n_{\rho },|m|+1,|m|+1,|m|+2,|m|+1,|m|+1,
\alpha_1^2\rho_f^2(z))] \nonumber \\
 & & -\frac{1}{2}[G(n_{\rho }',n_{\rho },|m|+1,|m|,\alpha _1^2\rho_i^2(z))
-G(n_{\rho }',n_{\rho},|m|+1,|m|,\alpha_1^2\rho_f^2(z))] \}
\nonumber \\
& & \cdot \langle n_z'|z-z_3|n_z \rangle _{z_i}^{z_f} \Big \}
\qquad ,m \ge 0
\end{eqnarray}

\begin{eqnarray}
\lefteqn{
\langle n_{\rho}'m'n_z'| \{ \mathbf \Omega _z(v_{g1}),
[\theta (z-z_i)-\theta(z-z_f)][\theta (\rho -\rho _i(z))-
\theta (\rho -\rho_f(z))] \} |n_{\rho}mn_z \rangle=} \nonumber \\
& & m_0\omega _g^2m\delta _{m'm}\{ [I(n_{\rho}',n_{\rho},|m|,|m|,|m|+1,
|m|+1,|m|,\alpha_1^2\rho_i^2(z)) \nonumber \\
 & & -I(n_{\rho }',n_{\rho },|m|,|m|,|m|+1,|m|+1,|m|,\alpha_1^2\rho_f^2(z))]
\nonumber \\
& & -\rho_3 \alpha _1 [I(n_{\rho}',n_{\rho},|m|,|m|,|m|+1,|m|+1,|m|-\frac{1}{2},
\alpha_1^2\rho_i^2(z)) \nonumber \\
 & & -I(n_{\rho}',n_{\rho},|m|,|m|,|m|+1,|m|+1,|m|-\frac{1}{2},
\alpha_1^2\rho_f^2(z)) ] \}
\end{eqnarray}

\newpage

\section*{FIGURE captions}

\bigskip \bigskip

\noindent {\bf   
Figure 1}  
Nuclear shape and deformation parameters for necked-in intersected ellipsoids.
$O_1$, $O_2$ and $O_3$ are the centers of the two deformed fragments and of
the neck sphere respectively. The four geometrical parameters which vary
are the two ratios of the ellipsoid semiaxes, $b_1/a_1$ and $b_2/a_2$, the
neck sphere radius $R_3$ and the distance between the fragment centers $R$.
\bigskip \bigskip

\noindent {\bf 
Figure 2}  
Sequences of shapes when geometrical parameters are varied. Neck sphere
radius vary from small (upper row - close to fusion shapes)
 through intermediary (second and third row)
up to large values for very elongated shapes. Along one row, the distance between 
centers is increased for the same value of the neck. Semiaxes ratios and 
mass asymmetry here are kept constant.

\bigskip \bigskip

\noindent {\bf
Figure 3}  
Matching potential ellipsoids: MPE between the two fragments (shaded
area) and between the neck potential and the fragment potentials (grey
area). The influence of each of the potentials is emphasized in the upper part.
The same figure with the surface functions for the regions of interest
(fragments: $\rho_1(z)$ and $\rho _2(z)$; neck: $\rho_g(z)$; MPE: $\rho(z)$;
MPE1: $\rho_{m1j}(z)$, $\rho _{m1s}(z)$, MPE2: $\rho_{m2s}(z)$ and $\rho_{m2j}(z)$
and the $z$-values used in calculations are marked. 

\bigskip \bigskip

\noindent {\bf
Figure 4}
The evolution  
the two fragment potentials $V_1(\rho,z)$, $V_2(\rho,z)$ and the neck
potential $V_g(\rho,z)$ for the $\rho$-value at the two tangent points of the
neck with the fragment ellipsoids: $\rho(z_{c1})$ - upper figure
and $\rho(z_{c2})$ when the distance between centers increases. One can observe
how the three potentials are tangent all along the splitting.

\bigskip \bigskip

\noindent {\bf Figure 5}  
Variation of the heavy fragment potential $V_1(\rho,z)$ and neck potential
$V_g(\rho,z)$ along the $O_1O_3$ direction. One can observe that inside
the neck sphere - hence inside MPE1 and MPE2, the neck potential is higher,
acting as a pressure element on the nuclear shape. The two crossing points
at $2R_3$ distance on $O_1O_3$ direction mark the limits of MPE1
along the $O_1O_3$ direction. 

\bigskip \bigskip

\noindent {\bf Figure 6}
Shell corrections calculated for the levels presented in Fig. 6, as a function of
the distance between centers $R$. The ground state is practically common.
A first bump appears in all cases; then for larger $R_3$ there is a decrease  
around $R$=10 fm. The second bump, which usualy generates a second maximum
in the fission barrier, lies around 15 fm, but is lower for small $R_3$.

\bigskip \bigskip

\noindent {\bf Figure 7}
Two level spectra with different mass asymmetry from the same superheavy parent
$^{306}122$ are presented on the upper part, for $R_3$=4 fm. 
The lower part of the figure shows
the corresponding shell correction energy
$E_{shell}$ as a function of $R$. First bump is present in both cases, eventhough
a little shifted. At the end of the splitting, asymmetric channel seems more
favorable from microscopic point of view, as shell correction energy is
some 4 MeV lower.

\bigskip \bigskip

\noindent {\bf Figure 8}
The level schemes for two fission channels from $^{252}$Cf are displayed
in the upper part of the figure. Levels are almost likely
in the first part of the splitting as can be deduced from the corresponding 
shell corrections. When the two fragments continue to split, the individual
wells influence results in a rather large difference in shell corrections, beyond
$R$=7 fm. 

\newpage 
\includegraphics[width=16cm]{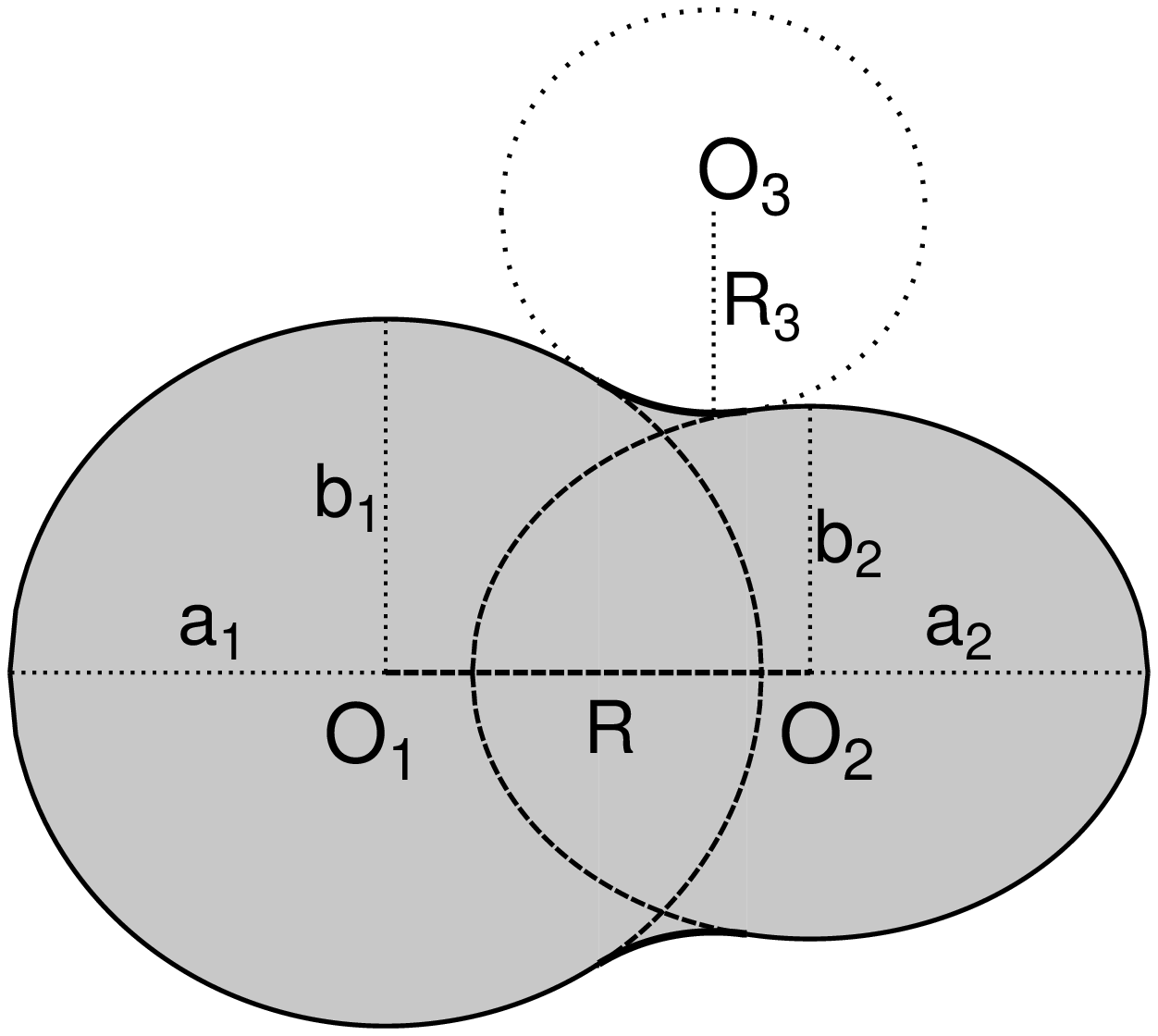}

\centerline{\bf Figure 1}

\bigskip

\includegraphics[width=16cm]{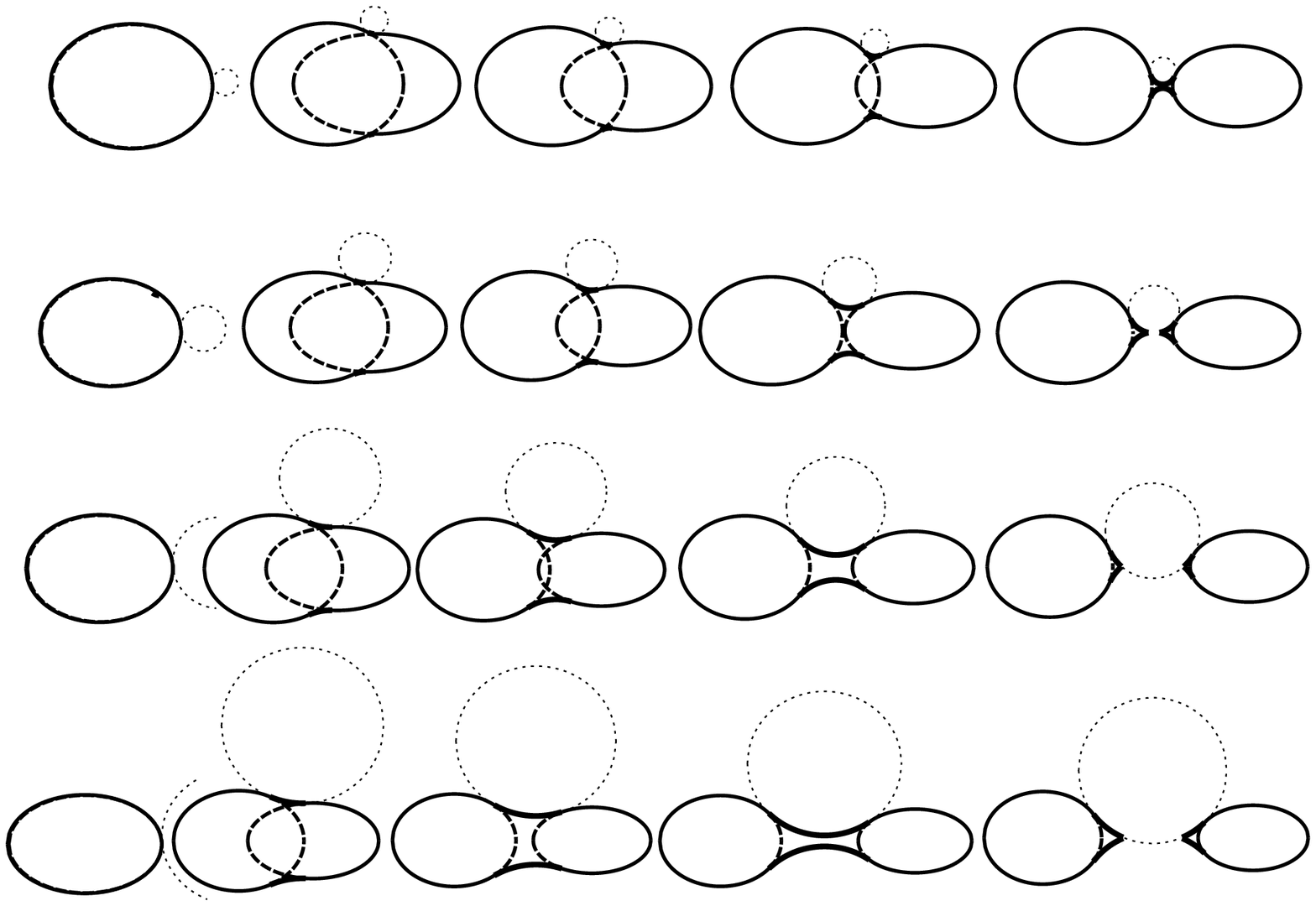}

\vspace*{-3cm}
\centerline{\bf Figure 2}


\includegraphics[width=16cm]{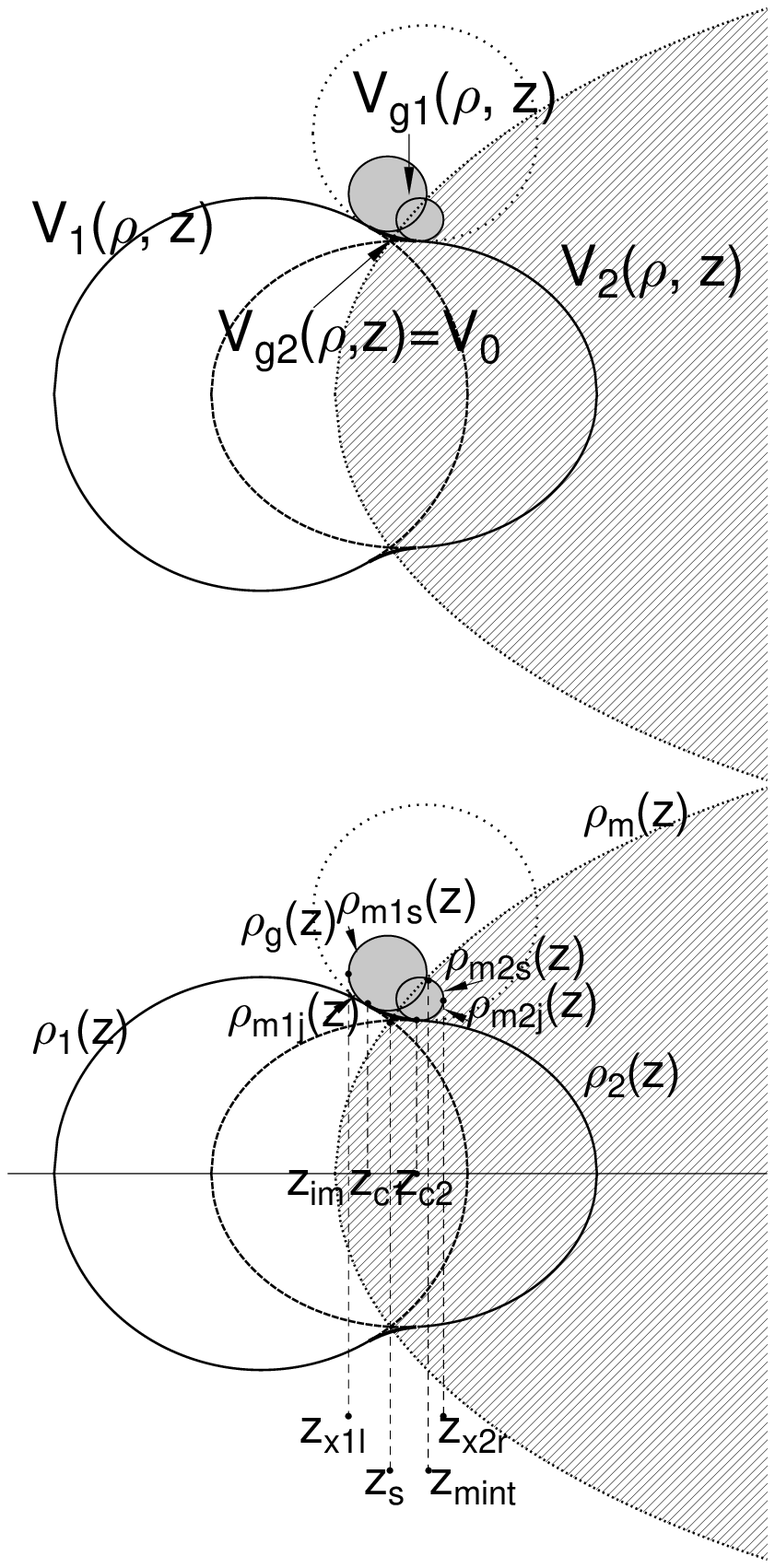}

\vspace{-2cm}
\centerline{\bf Figure 3}


\includegraphics[width=16cm]{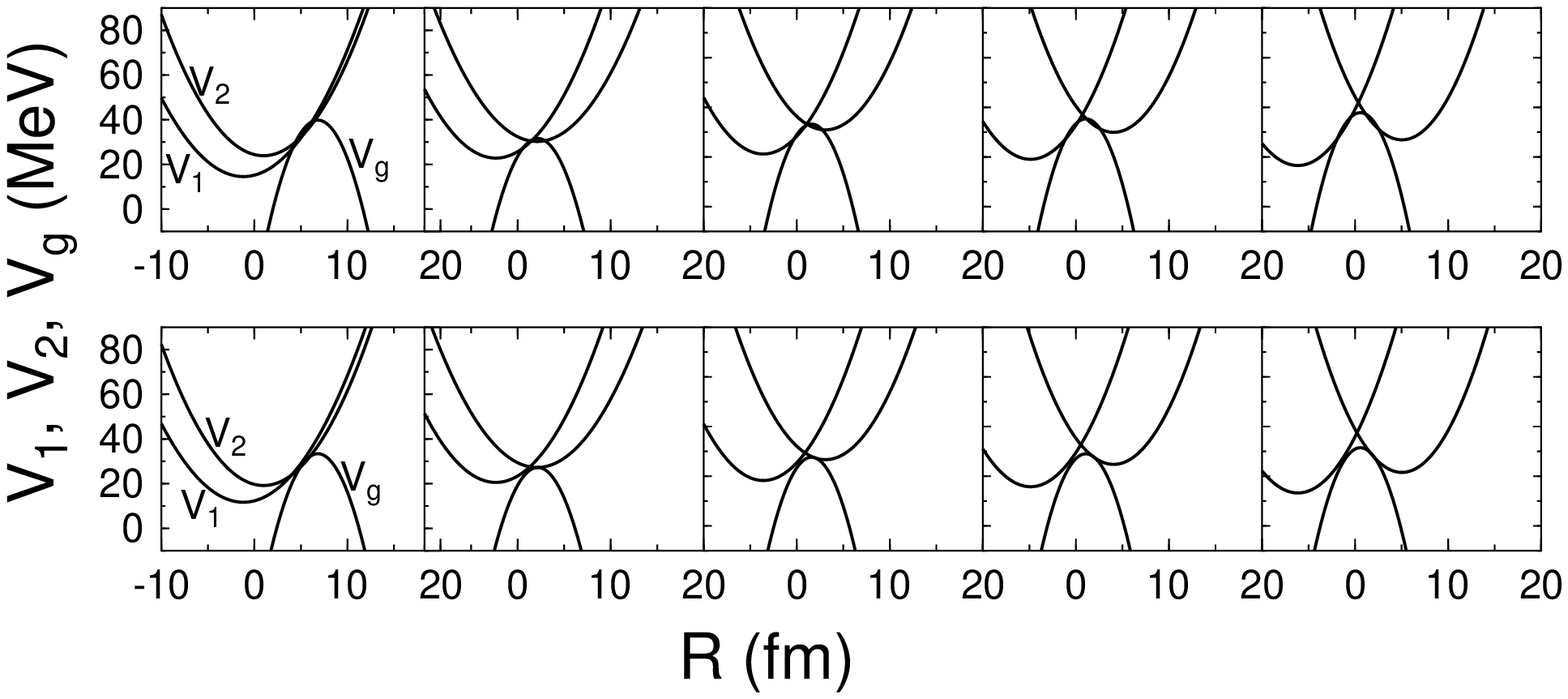}

\centerline{\bf Figure 4}


\includegraphics[width=16cm]{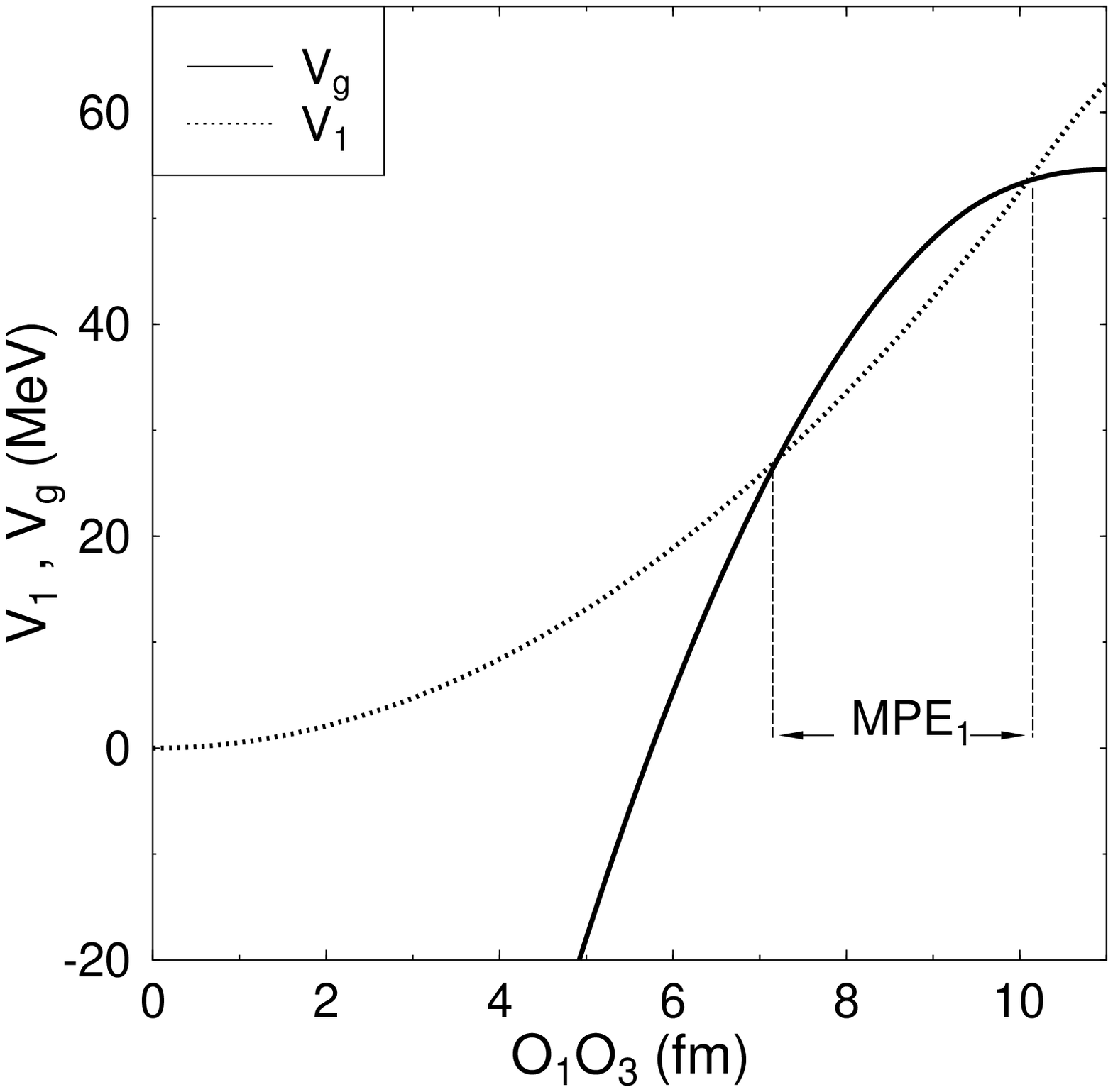}

\centerline{\bf Figure 5}

\includegraphics[width=16cm]{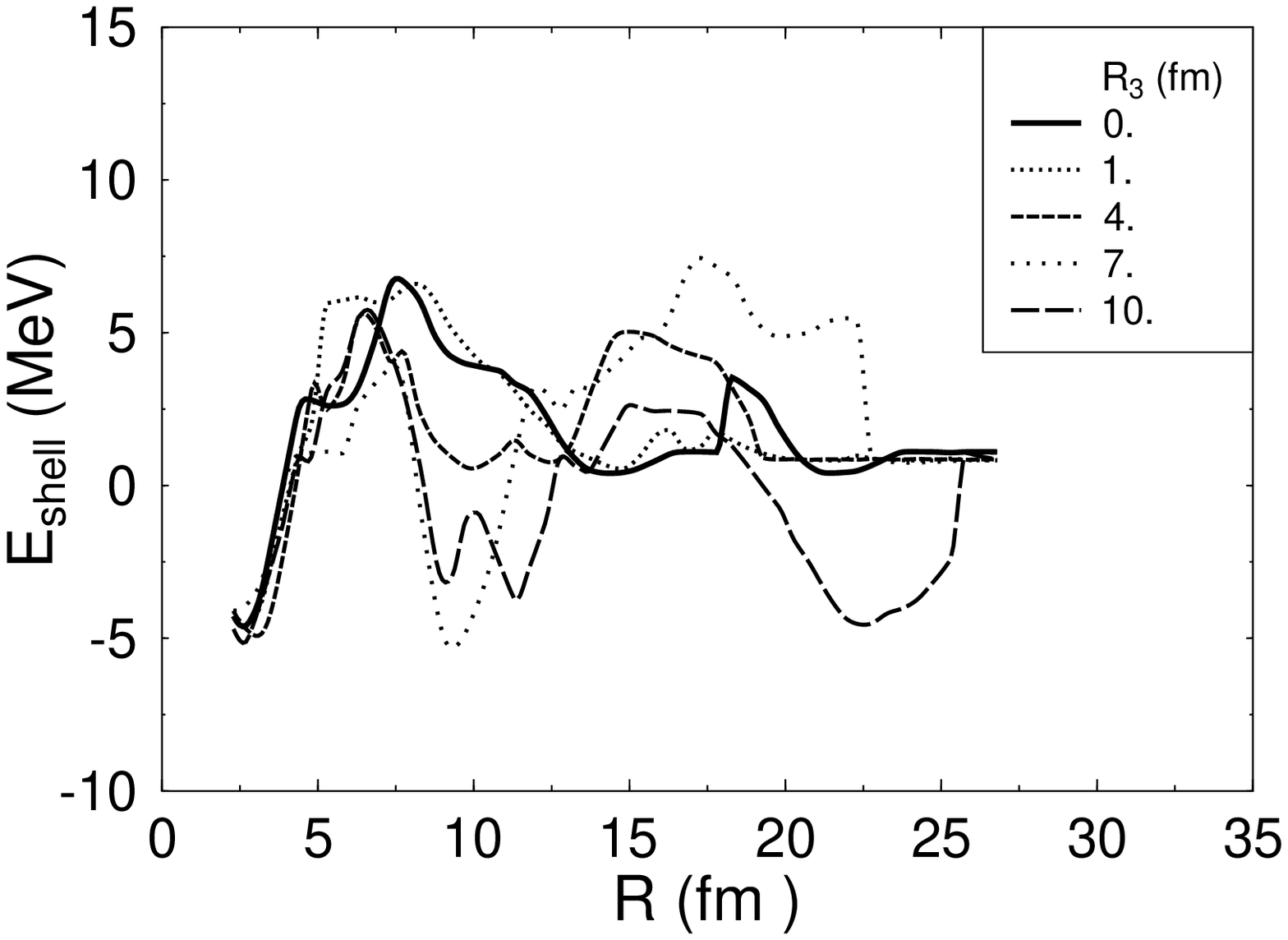}

\centerline{\bf Figure 6}

\includegraphics[width=16cm]{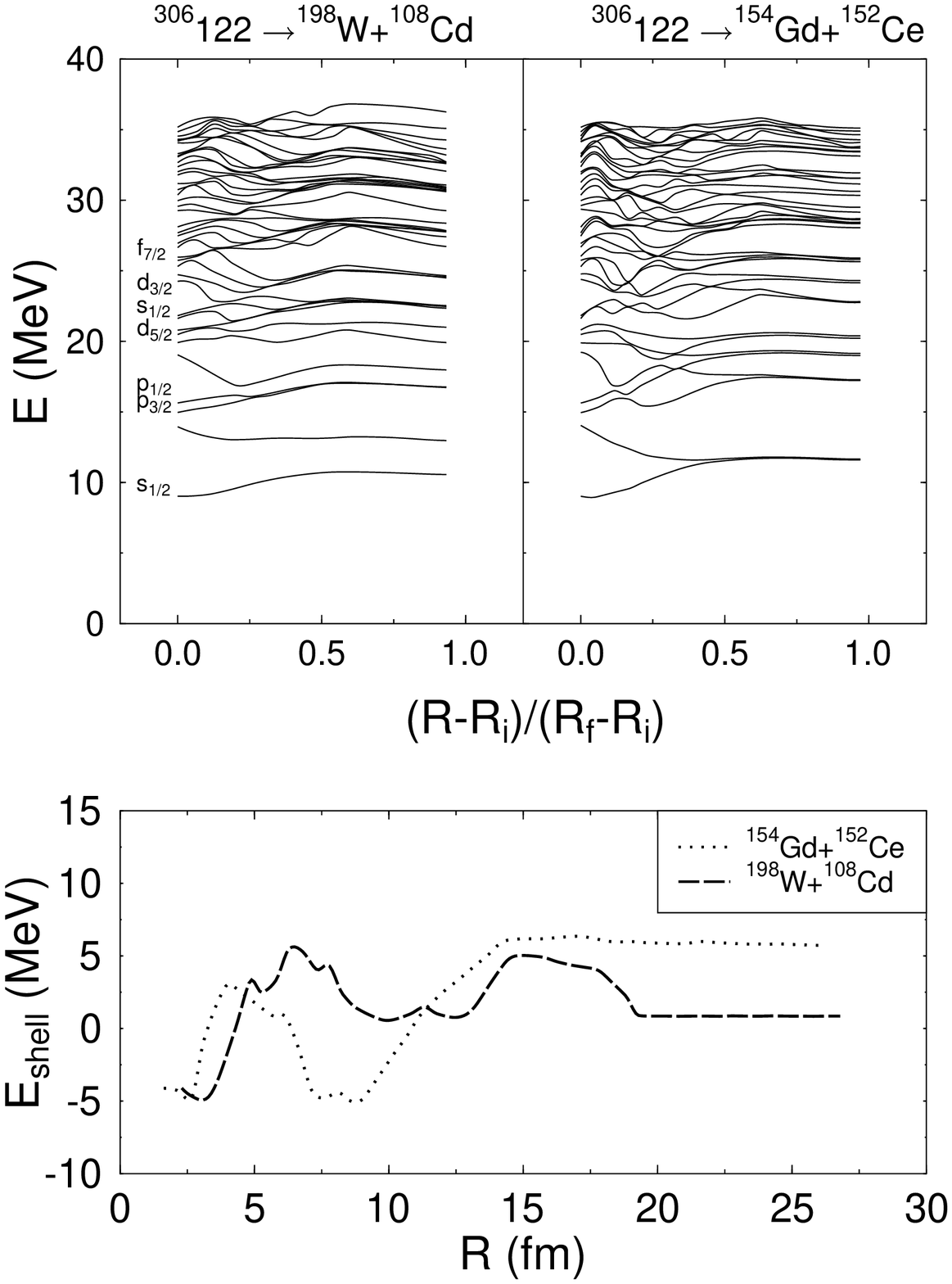}

\centerline{\bf Figure 7}

\includegraphics[width=16cm]{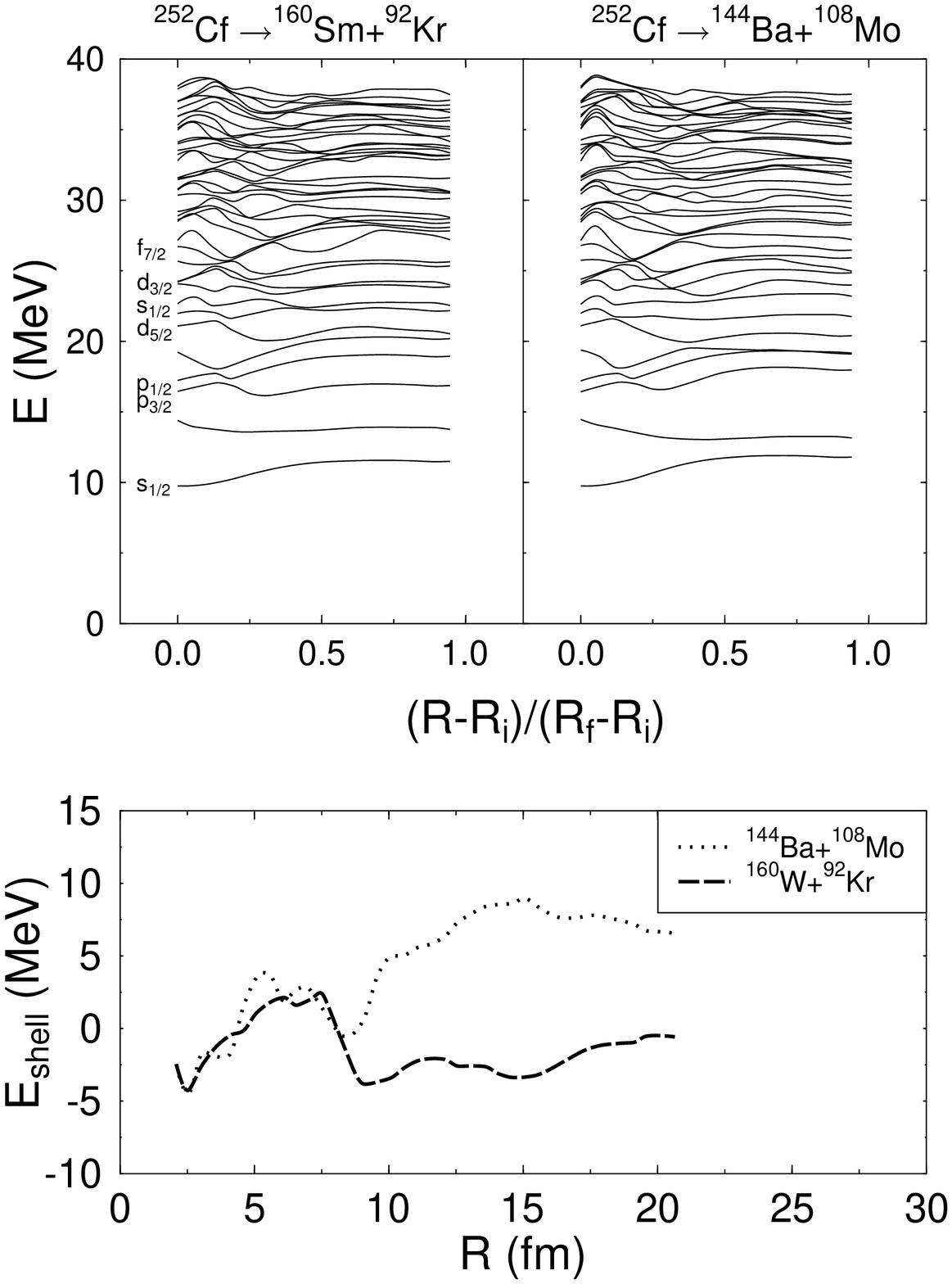}

\centerline{\bf Figure 8}

\end{document}